\documentclass[11pt]{article}
\usepackage{graphicx}
\usepackage[margin=1.25in]{geometry}
\usepackage[usenames,dvipsnames]{color}
\usepackage{url}
\usepackage[colorlinks = true,
            linkcolor = blue,
            urlcolor  = blue,
            citecolor = blue,
            anchorcolor = blue]{hyperref}

\usepackage{lineno,hyperref,color}
\modulolinenumbers[5]
\usepackage{todonotes}
\usepackage{booktabs}
\usepackage{float}
\usepackage{wrapfig}

\usepackage{appendix}
\usepackage{pdfpages}
\usepackage{hyperref}
\usepackage{geometry}                
\usepackage[parfill]{parskip}    
\usepackage{graphicx, subfigure}
\usepackage{amssymb}
\usepackage{amsmath}
\usepackage{epstopdf}
\usepackage{array}
\usepackage{lineno}
\usepackage{scrextend}

\def\Author#1{\begin{center}{ \sc #1} \end{center}}
\def\Address#1{\begin{center}{ \it #1} \end{center}}

\newenvironment{Abstract}{\begin{quotation} \begin{center}
                       ABSTRACT
     \end{center}\bigskip  }{\end{quotation}}

\def\beq{\begin{equation}}
\def\eeq#1{\label{#1}\end{equation}}
\def\eeqn{\end{equation}}

\newenvironment{Eqnarray}%
   {\arraycolsep 0.14em\begin{eqnarray}}{\end{eqnarray}}
\def\beqa{\begin{Eqnarray}}
\def\eeqa#1{\label{#1}\end{Eqnarray}}
\def\eeqan{\end{Eqnarray}}

\let\bar=\overbar

\def\lsim{\mathrel{\raise.3ex\hbox{$<$\kern-.75em\lower1ex\hbox{$\sim$}}}}
\def\gsim{\mathrel{\raise.3ex\hbox{$>$\kern-.75em\lower1ex\hbox{$\sim$}}}}

\def\del{\partial}
\def\Dslash{\not{\hbox{\kern-4pt $D$}}}
\def\dslash{\not{\hbox{\kern-2pt $\del$}}}
\def\pslash{\not{\hbox{\kern-2pt $p$}}}
\def\ETmiss{\not{\hbox{\kern-4pt $E$}}_T}

\def\Dlr{\mathrel{\raise1.5ex\hbox{$\leftrightarrow$\kern-1em\lower1.5ex\hbox{$D$}}}}

\def\MSB{{\bar{M \kern -2pt S}}}
\def\msb{{\bar{\scriptsize M \kern -1pt S}}}

\def\drb{{\bar{\scriptsize D \kern -1pt R}}}

\input{defs}

\newcommand\snowmass{\begin{center}\rule[-0.2in]{\hsize}{0.01in}\\\rule{\hsize}{0.01in}\\
\vskip 0.1in Submitted to the  Proceedings of the US Community Study\\ 
on the Future of Particle Physics (Snowmass 2021)\\ 
\rule{\hsize}{0.01in}\\\rule[+0.2in]{\hsize}{0.01in} \end{center}}

\begin{document}



\snowmass

{\bf\boldmath\LARGE
\begin{center}
Snowmass 2021 White Paper: \\
Charged Lepton Flavour Violation in Heavy Particle Decays
\end{center}
}


\bigskip 
\bigskip 

\Author{W.~Altmannshofer}
\Address{Department of Physics, University of California Santa Cruz, and
Santa Cruz Institute for Particle Physics, 1156 High St., Santa Cruz, CA 95064, USA}
\Author{C.~Caillol}
\Address{CERN, EP Department,
1 Esplanade des Particules, CH-1217 Meyrin, Switzerland}
\Author{M.~Dam\footnote{Corresponding author: dam@nbi.dk}, S.~Xella}
\Address{Niels Bohr Institute, Copenhagen University, Copenhagen, Denmark}
\Author{Y.~Zhang}
\Address{School of Physics, Southeast University, Nanjing 211189, China}

\medskip

\medskip

 \begin{Abstract}
\noindent Charged lepton flavor violation is an unambiguous signature for New Physics. Here we present a summary of the theoretical and experimental status of the search for charged lepton flavor violation in heavy particle decays, in particular in the decays of the Z and Higgs bosons, and of the top quark. Decays of beyond-Standard-Model particles such as a Z$'$ or an additional scalar particle are also discussed. Finally the prospects for such searches at proposed future electron-positron colliders are reviewed.
\end{Abstract}

\def\thefootnote{\fnsymbol{footnote}}
\setcounter{footnote}{0}
%


\section{Introduction}
\label{sec:introduction}

In the Standard Model (SM), with massless neutrinos, the three lepton-flavour quantum numbers are individually conserved. The discovery of neutrino oscillations demonstrate that neutrinos have mass and that lepton flavour can be violated in the neutral-lepton sector. Via loop diagrams, neutrino oscillations induce also lepton flavour violation (LFV) among charged leptons. These processes are however suppressed to an unobservable level by the tiny values of the neutrino masses. Any experimental observation of charged lepton flavour violation (CLFV) would therefore be an unambiguous signal for physics beyond the Standard Model (BSM) not related to the neutrino masses. 

Searches for CLFV processes can be divided into low- and high-energy categories. 
At low energy, the most sensitive probes are provided by the decays of $\mu$ and $\tau$ leptons and of K mesons, and by the $\mu \to \mbox{e}$ conversion in nuclei. At high energy, CLFV processes can be searched for both inclusively, in processes such as $\mbox{pp} \to \mu \text{e} + X$, and in the decays of heavy particles.
In this article we focus on the latter option and discuss the status of, and the future prospects for, the search for CLFV in the decays of the heaviest SM particles, the Z boson, the Higgs boson, and the top quark. Also a few BSM processes will be considered, in particular processes involving a possible Z$'$ boson or an additional scalar beyond the 125-GeV particle.

The paper is organised as follows. Section 2 starts out with a theoretical overview including the analysis of the sensitivity of CLFV decay processes to New Physics as described within the Standard Model Effective Field Theory framework. The discussion of CLFV processes involving BSM particles -- the Z$'$ and additional scalars -- will point also to how such phenomena could possibly explain the $g-2$ anomaly and how future collider experiments will be sensitive to the relevant regions of parameter space. 
Current results from the LHC and prospects from the 20-fold larger statistics from the HL-LHC are discusses in \mbox{Section 3}. 
Finally, Section 4 reviews the long-term prospects at future electron-positron colliders, where the clean experimental environments combined with very high luminosities, in particular at the proposed circular colliders, CEPC and FCC-ee, promise very sensitive CLFV tests.

\section{Theoretical Framework}
\label{sec:theory}

\subsection{Z Decays} \label{sec:theory:Zdecay}

As is the case with all LFV phenomena, SM predictions for the LFV Z decays, $\text{Z} \to \tau \mu$, $\text{Z} \to \tau e$, and $\text{Z} \to \mu e$, are suppressed by neutrino masses and thus completely beyond experimental reach. Taking into account neutrino masses, SM predictions for the branching fractions of such decays are in the ballpark of $10^{-50}$ to $10^{-40}$. Hence, any experimental evidence for LFV Z decays would be a clear sign of physics beyond the SM.

Direct searches for LFV Z decays into $\mu$e,  $\tau$e, and $\tau\mu$ final states were performed at LEP with branching fraction limits of about $2 \times 10^{-6}$ for the $\mu$e mode, and $10^{-5}$ for the $\tau$e and $\tau\mu$ modes~\cite{DELPHI97_CLFVZ,OPAL95_CLFVZ}. These limits have been recently improved
by measurements from ATLAS~\cite{ATLAS:2020zlz, ATLAS:2021bdj,ATLAS:2022uhq}, as detailed in \mbox{Sect.\ \ref{sec:lhcZ}}. Improvements were by about a factor of seven for the e$\mu$ mode and by factors of about two for the e$\tau$ and $\mu\tau$ modes. 
The new limits are
\begin{equation}
    \mathcal{B}(\text{Z} \to \mu\text{e})  < 2.62 \times 10^{-7}, \quad
    \mathcal{B}(\text{Z} \to \tau\text{e}) < 5.0 \times 10^{-6}, \quad
    \mathcal{B}(\text{Z} \to \tau\mu) < 6.5 \times 10^{-6}.
\end{equation}

The presence of new physics could increase the branching fraction expectations significantly bringing observations within experimental reach at the 
HL-LHC or future colliders. Among the many BSM scenarios that can give $\text{Z} \to \ell \ell^\prime$ decays are models with gauged flavor symmetries~\cite{Langacker:2000ju}, supersymmetric models~\cite{Brignole:2004ah}, models with heavy sterile neutrinos~\cite{Abada:2014cca,Abada:2015zea,DeRomeri:2016gum}, and leptoquark models~\cite{Crivellin:2020mjs}. 

New physics that leads to the LFV Z decays can be model independently parameterized in the context of the Standard Model Effectiv Field Theory (SMEFT)~\cite{Grzadkowski:2010es} which systematically extends the SM by higher dimensional interactions
\begin{equation}
 \mathcal L_\text{SMEFT} = \mathcal L_\text{SM} + \sum_{n > 4} \sum_{i_n} \frac{C_{i_n}}{\Lambda^{n-4}} Q_{i_n} ~.
\end{equation}
Here, $n$ denotes the mass dimension of the operators $Q_{i_n}$, $C_{i_n}$ are so-called Wilson coefficients, and $\Lambda$ is the new physics scale. Such a parameterization is valid as long as the new physics is heavy compared to the electro-weak scale.

The leading interactions that can give $\text{Z} \to \ell \ell^\prime$ decays at tree level are of dimension 6 and consist of dipole operators
\begin{equation}
 Q_{eW} = (\bar \ell_i \sigma^{\mu\nu} P_R \ell_j) \tau^I \phi W^I_{\mu\nu} ~,\quad Q_{eB} = (\bar \ell_i \sigma^{\mu\nu} P_R \ell_j) \phi B_{\mu\nu} ~,
\end{equation}
and Higgs current operators
\begin{eqnarray}
 Q_{\phi \ell}^{(1)} &=& (\phi^\dagger i \overset\leftrightarrow{D_\mu} \phi)(\bar \ell_i \gamma^\mu P_L \ell_j) ~,\quad Q_{\phi \ell}^{(3)} = (\phi^\dagger i \tau^I \overset\leftrightarrow{D_\mu} \phi)(\bar \ell_i \tau^I \gamma^\mu P_L \ell_j) ~, \nonumber \\
 \quad Q_{\phi e} &=& (\phi^\dagger i \overset\leftrightarrow{D_\mu} \phi)(\bar \ell_i \gamma^\mu P_R \ell_j) ~.
\end{eqnarray}
In the above operators, $i,j = 1,2,3$ label the lepton flavor, $\phi$ is the SM Higgs doublet, and $W_{\mu\nu}$ and $B_{\mu\nu}$ are the field strength tensors of the $SU(2)_L$ and $U(1)_Y$ gauge bosons. 
Assuming generic Wilson coefficients $C \sim 1$, one finds the following order of magnitude estimate for the branching ratios of flavor changing $Z$ decays
\begin{equation}
 \mathcal{B}(\text{Z} \to \ell \ell^\prime) \sim \left( \frac{v}{\Lambda} \right)^4 ~, 
\end{equation}
where $v = 246$\,GeV is the vacuum expectation value of the Higgs. The existing bounds thus already probe new physics at few TeV.

\begin{figure}[tb]
 \centering
  \includegraphics[width=1.0\textwidth]{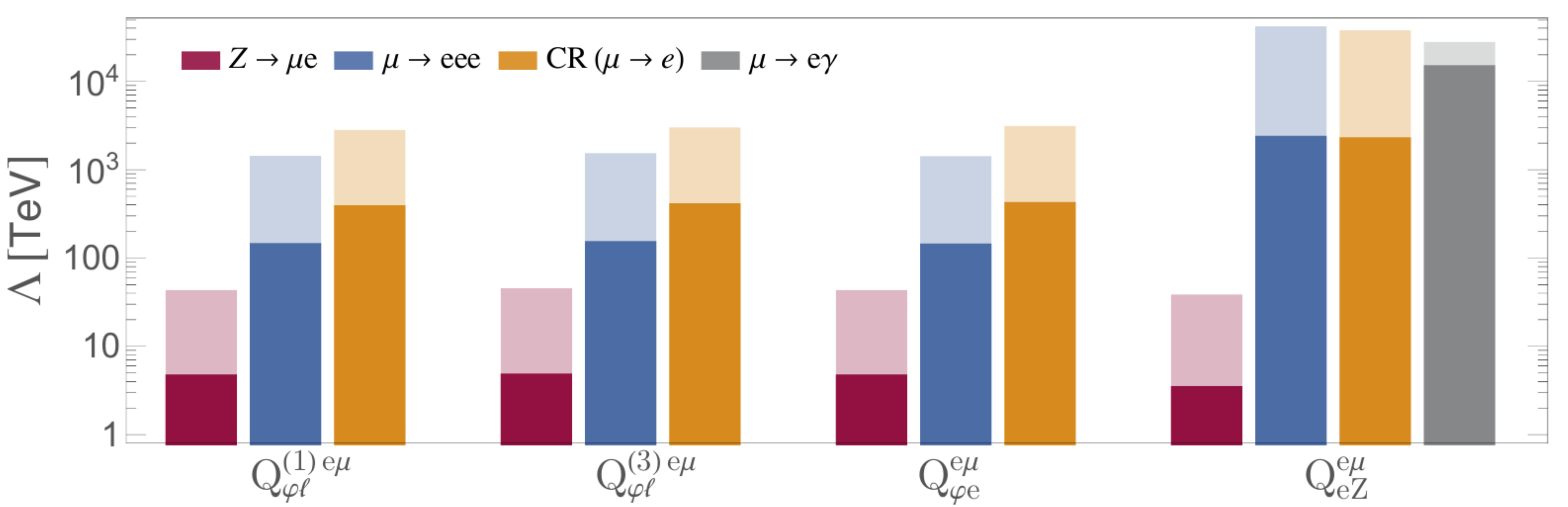}
  \vspace{-15pt}
  \caption{New physics sensitivity of the $\text{Z} \to \mu e$ decay compared to several low energy lepton flavor violating processes. One SMEFT operator is switched on at a time. Shown is the reach in the new physics scale setting the Wilson coefficient to 1. Dark bars correspond to the current status, the light bars show expected future sensitivities at a Tera-Z, MEG-II, Mu3e, and Mu2e/COMET (from~\cite{Calibbi:2021pyh}).}
  \label{fig:Zmue}
\end{figure}
\begin{figure}[tb]
 \centering
  \includegraphics[width=1.0\textwidth]{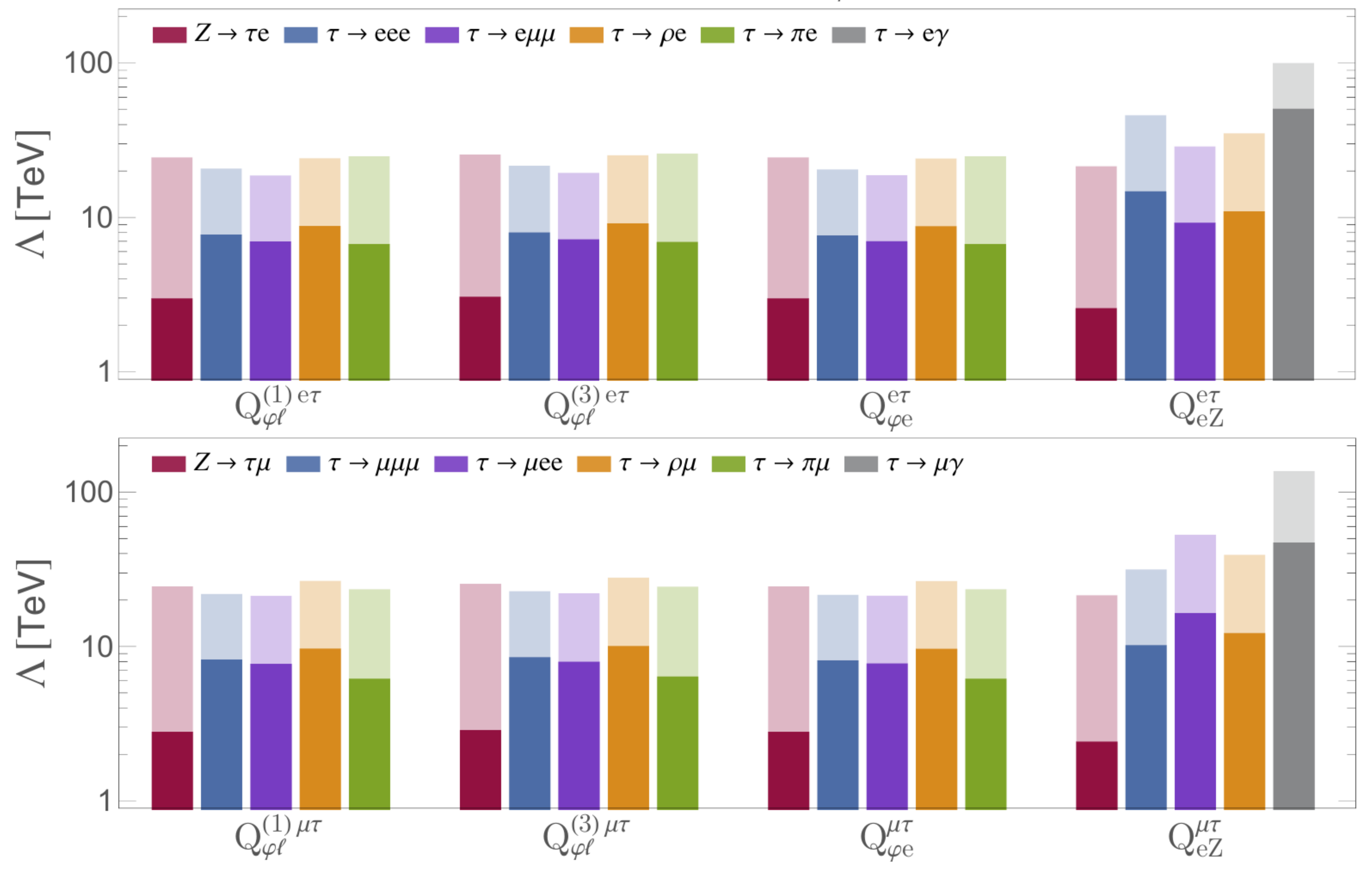}
  \vspace{-15pt}
  \caption{New physics sensitivity of the $\text{Z}\to \tau \mbox{e}$ and $\text{Z} \to \tau \mu$ decays compared to several low energy lepton flavor violating processes. One SMEFT operator is switched on at a time. Shown is the reach in the new physics scale setting the Wilson coefficient to 1. Dark bars correspond to the current status, the light bars show expected future sensitivities at a Tera-Z and Belle II (from~\cite{Calibbi:2021pyh}).}
  \label{fig:Ztaumu_taue}
\end{figure}

It is important to note that in the presence of the effective operators listed above, one not only expects LFV Z decays, but at the same time also various LFV phenomena at low energies, including radiative, leptonic, and semi-leptonic decays of muons and taus, e.g. $\mu \to \mbox{e} \gamma$, $\tau \to \mu \gamma$, $\tau \to 3$e, $\tau \to \mu \pi^0$, etc. and $\mu \to \mbox{e}$ conversion in nuclei. Barring accidental cancellations of the new physics in the low energy processes, LFV muon decays $\mu \to \mbox{e} \gamma$, $\mu \to \mbox{3e}$, and $\mu \to \mbox{e}$ conversion in nuclei outperform $\text{Z} \to \mu \mbox{e}$ decays in terms of new physics sensitivity. This is illustrated in \mbox{Fig.\ \ref{fig:Zmue}} that shows the maximal new physics scale that can be probed with current and future experimental sensitivities.

A different picture emerges for LFV processes involving taus. In this case, the $\text{Z}\to \tau \mbox{e}$ and $\text{Z} \to \tau \mu$ decays have sensitivities to new physics that are comparable to tau decays, see \mbox{Fig.\ \ref{fig:Ztaumu_taue}}. The Z decays and the $\tau$ decays are thus highly complementary in probing new physics~\cite{Calibbi:2021pyh,Davidson:2012wn}.


\subsection{Higgs Decays}  \label{sec:theory:Hdecay}

SM predictions for LFV Higgs decays, $h \to \tau \mu$, $h\to \tau \mbox{e}$, and $h\to \mu \mbox{e}$, are likewise suppressed by neutrino masses, and completely beyond experimental reach. Any observation of such processes would be a clear signature for new physics.

To provide a phenomenological parameterization of possible BSM effects in Higgs decays, it is useful to introduce the following flavor violating Higgs couplings in the fermion mass eigenstate basis
\begin{equation}
 \mathcal{L}_\text{pheno} \supset - \sum_{i\neq j = 1,2,3} \left[\left( \kappa_{\ell_i \ell_j} + i \tilde \kappa_{\ell_i \ell_j} \right) h \bar \ell_i P_R \ell_j  ~+~\text{h.c.} \right] ~,
\end{equation}
where $i,j = 1,2,3$ label the three generations of leptons. The coefficients $\kappa$ and $\tilde \kappa$ are the real and imaginary parts of the couplings.
In terms of these phenomenological couplings, branching ratios of flavor violating Higgs decays are given by
\begin{equation}
\mathcal{B}(h \to \ell_i \ell_j) = \frac{m_h}{8\pi \Gamma_h} \left( \kappa_{\ell_i \ell_j}^2 + \tilde \kappa_{\ell_i \ell_j}^2 + \kappa_{\ell_j \ell_i}^2 + \tilde \kappa_{\ell_j \ell_i}^2 \right) ~,
\end{equation}
where $\Gamma_h$ is the total Higgs width.

There are indirect constraints on LFV Higgs couplings. The strongest ones are from radiative muon and tau decays $\mu \to \mbox{e} \gamma$, $\tau\to\mu\gamma$, and $\tau\to \mbox{e} \gamma$. Using existing bounds on the muon and tau decays, one finds (adapted from~\cite{Harnik:2012pb}, see also \cite{Blankenburg:2012ex})
\begin{equation}
 \mathcal{B}(h \to \tau \mu)_\text{ind.} \lesssim 0.32 ~, \qquad
 \mathcal{B}(h \to \tau \mbox{e})_\text{ind.} \lesssim 0.24 ~,\qquad 
 \mathcal{B}(h \to \mu \mbox{e})_\text{ind.} \lesssim 1.6 \times 10^{-8} ~.
\end{equation}
Therefore, observing $h \to \mu \mbox{e}$ in the foreseeable future would not only imply new physics, but at the same time also requires a non-trivial mechanism to suppress $\mu \to \mbox{e} \gamma$ below existing constraints.
In the following, we thus focus on LFV Higgs decays involving taus which could be sizable.

Direct searches for the decays $h \to \tau \mu, \tau e$ are carried out at the LHC~\cite{ATLAS:2019pmk, CMS:2021rsq}. Existing limits are
\begin{eqnarray}
 &&\mathcal{B}(h \to \tau \mu)_\text{ATLAS} \lesssim 2.8 \times 10^{-3}~, \qquad
 \mathcal{B}(h \to \tau \mu)_\text{CMS} \lesssim 1.5 \times 10^{-3}~, \\
 &&\mathcal{B}(h \to \tau \mbox{e})_\text{ATLAS} \lesssim 4.7 \times 10^{-3} ~,\ \qquad
 \mathcal{B}(h \to \tau \mbox{e})_\text{CMS} \lesssim 2.2 \times 10^{-3}~.
\end{eqnarray}
Sensitivities will improve considerably at the high luminosity LHC and at future colliders (see Sects.\ \ref{sec:lhc} and~\ref{sec:ee}).

\begin{figure}[tbh]
 \centering
  \includegraphics[width=0.48\textwidth]{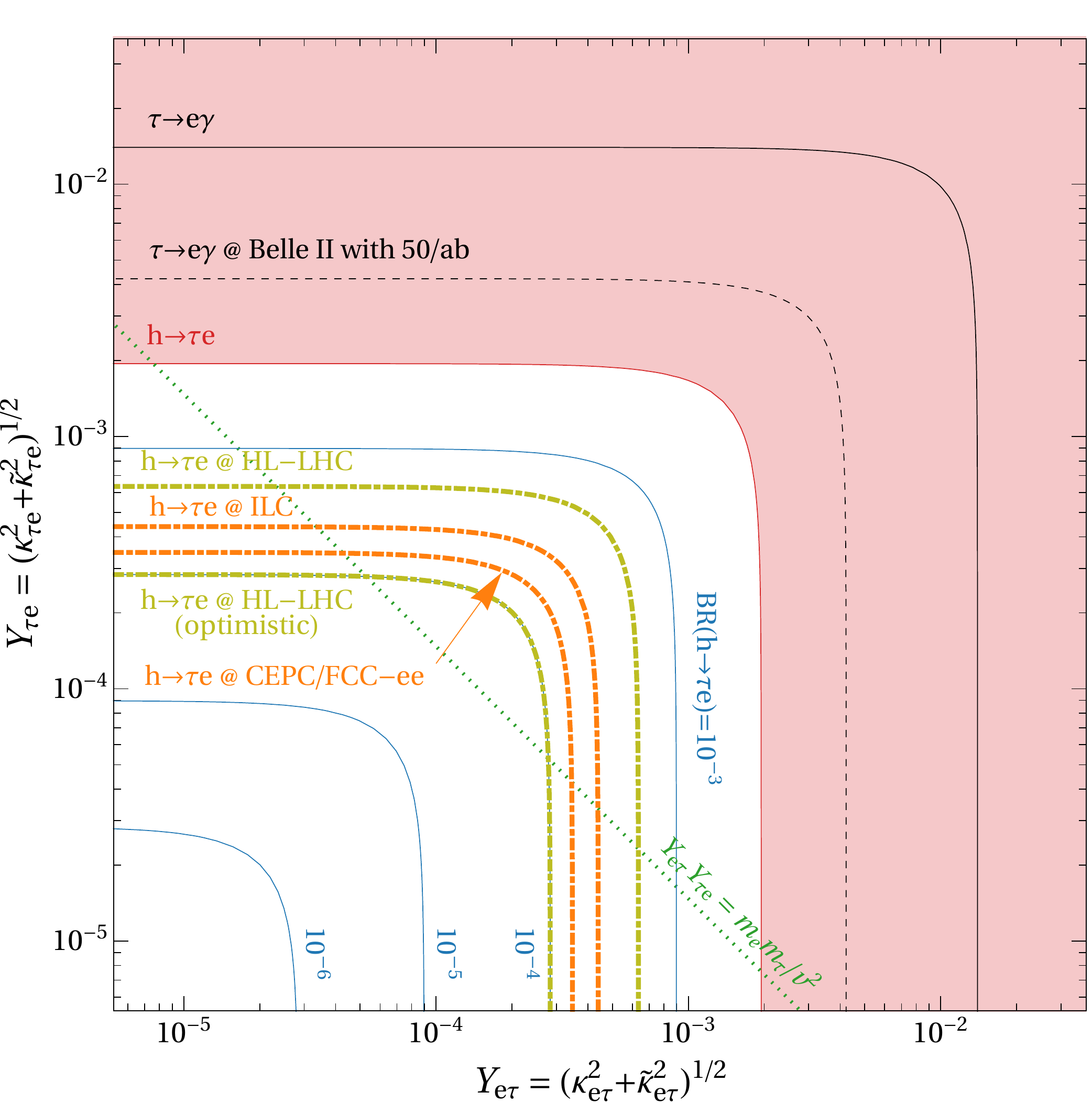} ~
  \includegraphics[width=0.48\textwidth]{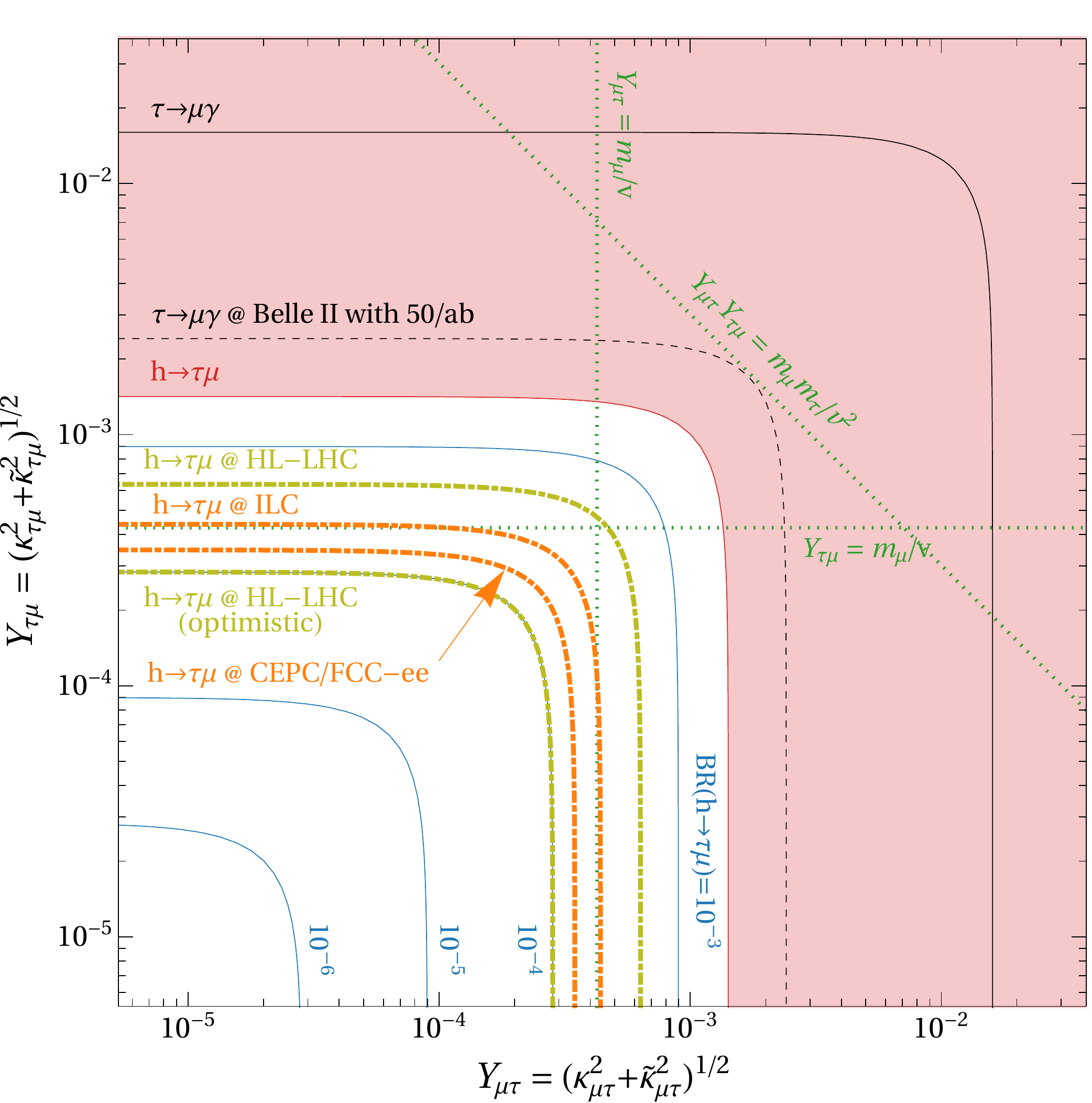}
  \caption{Current constraints (red shaded regions) and expected sensitivities (orange and yellow dash-dotted lines) in the plane of lepton flavor violating Higgs couplings. The solid blue lines show the predicted values for $\mathcal{B}(h \to \tau \mu)$ and $\mathcal{B}(h \to \tau e)$. The dotted green lines indicate theoretically motivated targets for the Higgs couplings.}
  \label{fig:Ytaumu_Ytaue}
\end{figure}

The plots in Fig.~\ref{fig:Ytaumu_Ytaue} show current constraints (red shaded regions) and expected sensitivities (orange and yellow dash-dotted lines) in the plane of LFV Higgs couplings. The predicted values for $\mathcal{B}(h \to \tau \mu)$ and $\mathcal{B}(h \to \tau e)$ are shown by the solid blue lines. Theoretically motivated targets for the Higgs couplings are indicated by the dotted green lines. 

Assuming that the new physics is sufficiently heavy compared to the electro-weak scale $v = 246$\,GeV, the LFV Higgs couplings are expected to scale as $\kappa_{\ell_i \ell_j}, \tilde \kappa_{\ell_i \ell_j} \propto \frac{v^2}{\Lambda^2}$ where $\Lambda$ is the scale of new physics. 
In the context of SMEFT~\cite{Grzadkowski:2010es}, the following dimension six operators lead to the flavor violating couplings of the Higgs
\begin{equation}
 \mathcal L_\text{SMEFT} \supset \sum_{i,j = 1,2,3} \frac{C_{ij}}{\Lambda^2} \left[\bar \ell_i P_R \ell_j \phi  ~+~\text{h.c.} \right] \phi^\dagger \phi \quad \Rightarrow \quad  \kappa_{\ell_i \ell_j} + i \tilde \kappa_{\ell_i \ell_j} = \frac{C_{ij}}{\sqrt{2}} \frac{v^2}{\Lambda^2} ~,
\end{equation}
where $\phi$ is the SM Higgs doublet. 
For generic Wilson coefficients $C_{ij} \sim 1$, sensitivty to branching ratios of $\mathcal O(10^{-4})$ probes new physics scales of $\Lambda \sim 10$\,TeV.

Many new physics models can lead to LFV Higgs couplings. They arise in models with heavy vector-like fermions or lepto-quarks, or in models with a new source of electro-weak symmetry breaking (e.g. a second Higgs doublet or a technifermion condensate).
The simplest models with vector-like leptons or lepto-quarks are thightly constrained by LFV tau decays~\cite{Falkowski:2013jya, Dorsner:2015mja}. Branching ratios of $h \to \tau\mu$ and $h \to \tau e$ at an observable level cannot be accommodated in these models. EFT arguments suggest this is a generic feature in models without new sources of electro-weak symmetry breaking~\cite{Altmannshofer:2015esa} (see however the exceptions discussed in~\cite{Altmannshofer:2016oaq, Galon:2017qes}).

If a new physics model contains a new source of electro-weak symmetry breaking, branching ratios of LFV Higgs decays can be sizable. Models with extended Higgs sectors have been extensively studied in this respect~\cite{Davidson:2010xv, Dery:2014kxa, Campos:2014zaa, AristizabalSierra:2014trk, Heeck:2014qea, Omura:2015nja, Crivellin:2015mga, Altmannshofer:2015esa, Botella:2015hoa, Aloni:2015wvn, Huitu:2016pwk, Altmannshofer:2016zrn, Hou:2020tgl}. It has been found that the $h\to\tau\mu$ and $h \to \tau \mbox{e}$ branching ratios can saturate the current direct bounds in many of such models.


\subsection{Top Decays}  \label{sec:theory:tdecay}

In contrast to the LFV decays of the Z and Higgs, LFV decays of the top quark are necessarily 3-body decays, $t \to q \ell \ell^\prime$, $q = u,c$. As these 3-body decays have to compete with a leading unsuppressed 2-body decay mode, $t \to b W$, their generic new physics reach is slightly lower. However, the top decays probe qualitatively different types of new physics and are thus highly complementary to the Z and Higgs decays discussed above. 

As is the case for the Z and Higgs decays, new physics that gives rise to the LFV top decays can be systematically and model independently parameterized by effective dimension 6 interactions of the SMEFT. The relevant operators are~\cite{Grzadkowski:2010es,Davidson:2015zza}
\begin{eqnarray} \label{eq:theory:raret1}
 Q_{\ell q}^{(1)} = (\bar \ell_i \gamma^\mu P_L \ell_j)(\bar q \gamma_\mu P_L t) ~&,&~~~  Q_{\ell q}^{(3)} = (\bar \ell_i \gamma^\mu \tau^I P_L \ell_j)(\bar q \gamma_\mu \tau^I P_L t) ~, \\  \label{eq:theory:raret2}
 Q_{\ell u} = (\bar \ell_i \gamma^\mu P_L \ell_j)(\bar q \gamma_\mu P_R t) ~&,&~~~  Q_{qe} = (\bar \ell_i \gamma^\mu P_R \ell_j)(\bar q \gamma_\mu P_L t) ~, \\  \label{eq:theory:raret3}
 Q_{eu} = (\bar \ell_i \gamma^\mu P_R \ell_j)(\bar q \gamma_\mu P_R t) ~&,& \\  \label{eq:theory:raret4}
 Q_{\ell e qu}^{(1)} = (\bar \ell_i P_R \ell_j)(\bar q P_R t) ~&,&~~~  Q_{\ell e qu}^{(3)}  = (\bar \ell_i \sigma_{\mu\nu} P_R \ell_j)(\bar q \sigma^{\mu\nu} P_R t) ~,
\end{eqnarray}
with $q = u,c$. 

Assuming that the Wilson coefficients of these operators are of $\mathcal O(1)$, the generic estimate for the top decay branching ratios is
\begin{equation}
 \mathcal{B}(t \to q \ell \ell^\prime) \sim \frac{1}{16\pi^2} \left( \frac{v}{\Lambda} \right)^4 ~, 
\end{equation}
with the Higgs vev $v$ and the scale of new physics $\Lambda$.

More than $10^8$ top quarks were produced at Run\,2 of the LHC, opening up the possibility to search for rare LFV top quark decays with high sensitivity. A recent CMS analysis~\cite{CMS:2022ztx} finds bounds on the $t \to u \mu e$ branching ratio of approximately $10^{-7}$ and on the $t \to c \mu e$ branching ratio of approximately $10^{-6}$. The analysis, discussed in more detail in Sect.\ \ref{sec:lhcTop}, includes both searches for top decays as well as searches for single top production in association with $\mu e$. Due to differences in the experimental acceptances, the constraints differ by an $\mathcal O(1)$ amount for the vector, scalar, and tensor operators in~\eqref{eq:theory:raret1}--\eqref{eq:theory:raret4}. The bounds obtained by CMS already probe new physics above 1~TeV. 

Some of the operators that give the $t \to q \ell \ell^\prime$ decays can also be probed by different processes. In particular, the operators containing left-handed charm or up quarks can lead to sizable rates of LFV kaon or $B$ meson decays like $K_L \to \mu e$, $B \to \pi \ell \ell^\prime$, or $B \to K \ell \ell^\prime$~\cite{Davidson:2015zza}. However, for operators with right-handed up and charm quarks, the indirect constraints are very weak and the LFV top decays provide the most sensitive probes.


\subsection{Z$'$ Boson}

A heavy or light Z$'$ boson exists in a large variety of new physics models, and the corresponding phenomenologies are very rich; see e.g.\ Refs.~\cite{Langacker:2000ju, Langacker:2008yv, delAguila:2010mx}. The LFV couplings of the Z$'$ boson can originate from the mixing of SM fermions with heavy exotic fermions~\cite{Langacker:1988ur}. The muon $g-2$ anomaly is an important motivation for LFV couplings of a Z$'$ boson~\cite{Lindner:2016bgg}. Without loss of generality, the effective LFV couplings of the Z$'$ can be written as, 
\begin{eqnarray}
\label{eqn:Zprime}
{\cal L} \supset Z_\mu^\prime \left( g'_{L} \bar{\ell}_{\alpha\,L} \gamma^\mu \ell_{\beta\,L} + g'_{R} \bar{\ell}_{\alpha\,R} \gamma^\mu \ell_{\beta\,R} \right) ~+~ {\rm H.c.} \quad (\alpha \neq \beta) \,.
\end{eqnarray}
Note that the couplings of the Z$'$ to left- and right-handed charged leptons can be different. Direct searches of a heavy Z$' \to \text{e}\mu,\, \text{e}\tau,\, \mu\tau$ have been performed at the LHC~\cite{CMS:2022fsw}. Assuming the LFV couplings of the Z$'$ are the same as the couplings of the Z boson to charged leptons, Z$'$ masses have been excluded up to 5.0\,TeV in the e$\mu$ channel, 4.3\,TeV in the e$\tau$ channel, and 4.1\,TeV in the $\mu\tau$ channel~\cite{CMS:2022fsw}. 

\begin{figure}[!t]
  \centering
  \includegraphics[width=0.55\textwidth]{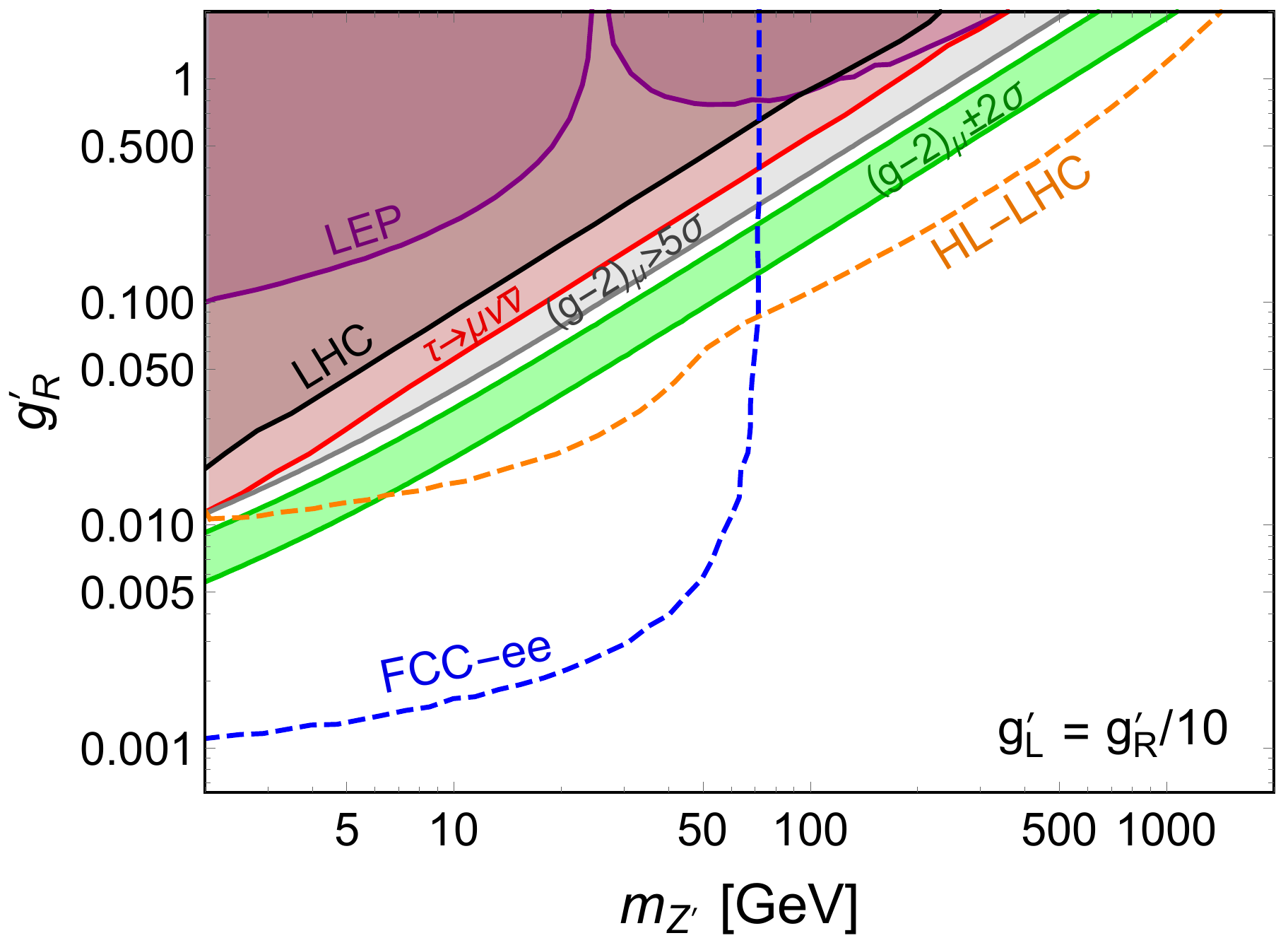} 
  \caption{Prospect of $m_\mathrm{Z'}$ and $g'_R$ at HL-LHC (dashed orange line) and FCC-ee (dashed blue line) with $g'_L = g'_R/10$. The green band correspond to the muon $g-2$ anomaly at the $2\sigma$ CL. Also shown are the limits from the muon $g-2$ discrepancy at the $5\sigma$ CL (gray), $\tau \to \mu \nu\bar{\nu}$ (red), LHC W data (black) and the LEP Z-pole data (purple). Figure remade based on Ref.~\cite{Altmannshofer:2016brv}. }
  \label{fig:Zprime}
\end{figure}
The LFV coupling $g'_{L,\,R}$ of the light Z$'$ boson can be directly measured at the high-energy hadron and lepton colliders via the process 
\begin{eqnarray}
\label{eqn:Zprime2}
pp,\, \text{e}^+ \text{e}^- \to \ell_\alpha^\pm \ell_\beta^\mp \text{Z}' \,.
\end{eqnarray}
Take the explicit example of $\alpha = \mu$ and $\beta = \tau$, with the decay Z$' \to \mu^\pm \tau^\mp$, the LFV coupling $g'_{L,\,R}$ will generate same-sign dilepton pairs  $\mu^\pm \mu^\pm \tau^\mp \tau^\mp$. With $g'_L = g'_R/10$, the prospects of Z$'$ mass $m_\mathrm{Z'}$ and $g'_R$ at the HL-LHC 14 TeV with an integrated luminosity of 3 ab$^{-1}$ and the FCC-ee 91 GeV with 150 ab$^{-1}$ are presented respectively as the dashed orange and blue curves in Fig.~\ref{fig:Zprime}~\cite{Altmannshofer:2016brv}. At HL-LHC, the Z$'$ boson can be probed up to the TeV scale; at the future lepton collider FCC-ee, the coupling $g'_R$ can be probed down to $10^{-3}$.

For low Z$'$ masses, $m_\mathrm{Z'} \lesssim m_\tau - m_\mu$, 
the Z$'$ boson can be produced from the two-body tau decay $\tau \to \mu \text{Z}'$ via the gauge couplings in Eq.~(\ref{eqn:Zprime}). Searches for two-body decay $\tau \to \mu +$inv.\ have been performed by ARGUS, and the null results exclude coupling $g'_R \gtrsim 10^{-6}$~\cite{ARGUS:1995bjh}. For higher masses, $m_\mathrm{Z'} \gtrsim m_\tau - m_\mu$, the coupling $g'_R$ is constrained by lepton flavor universality results in tau decays, high-precision LHC W data, LEP Z-pole data, and the muon $g-2$ anomaly. These limits are shown,  at $5\sigma$ CL, as the shaded regions in Fig.~\ref{fig:Zprime}. With $\tau$ running in the loop, LFV couplings of a Z$'$ boson can explain the muon $g-2$ discrepancy, which is in some sense similar to the neutral scalar case described below~\cite{Lindner:2016bgg, Altmannshofer:2016brv}. The corresponding $2\sigma$ regions for the muon $g-2$ anomaly is shown as the green band in Fig.~\ref{fig:Zprime}. As seen in Fig.~\ref{fig:Zprime}, when all these constraints are taken into consideration, a Z$'$ with mass $m_\mathrm{Z'} \gtrsim 2$\,GeV can provide a viable interpretation of the muon $g-2$ anomaly. Furthermore, both the future high-energy hadron and lepton colliders can probe large regions of the Z$'$ interpretation of muon $g-2$ anomaly. 

With LFV couplings to electron and muon, the Z$'$ boson could contribute to $\mu -\text{e}$ scattering~\cite{Dev:2020drf, Masiero:2020vxk}, e.g.\ in the MUonE experiment, which is proposed to determine the contribution of hadronic vacuum polarization to muon $g-2$~\cite{CarloniCalame:2015obs, Abbiendi:2016xup}. However, the LFV couplings of Z$'$ boson is tightly constrained by muonium-antimuonium oscillation and electron $g-2$. These limits have precluded the sensitivities of the MUonE experiment to the lepton flavor violating couplings of Z$'$ boson~\cite{Dev:2020drf}.

\subsection{BSM Neutral Scalar}  \label{sec:theory:NScalar}

%
In a large variety of BSM scenarios, there exist BSM neutral scalars $H$, which couple to the SM charged leptons in a flavor-violating way at the tree- or 1-loop level~\cite{Hou:1995dg, Dev:2017ftk, Li:2018cod, Arganda:2019gnv}.
%
%
In the perspective of effective theory, the couplings of $H$ to the SM charged leptons can be written in the following way, without loss of generality,
\begin{equation}
    {\cal L}_Y = h_{\alpha\beta} \overline{\ell}_\alpha H \ell_\beta ~+~ {\rm H.c.} \,,
\end{equation}
where 
$h_{\alpha\beta}$ is the Yukawa coupling. For simplicity, we have assumed $H$ to be a real field and CP-even. As a result, the matrix $h_{\alpha\beta}$ is symmetric.

\begin{figure}[!t]
  \centering
  \includegraphics[width=0.32\textwidth]{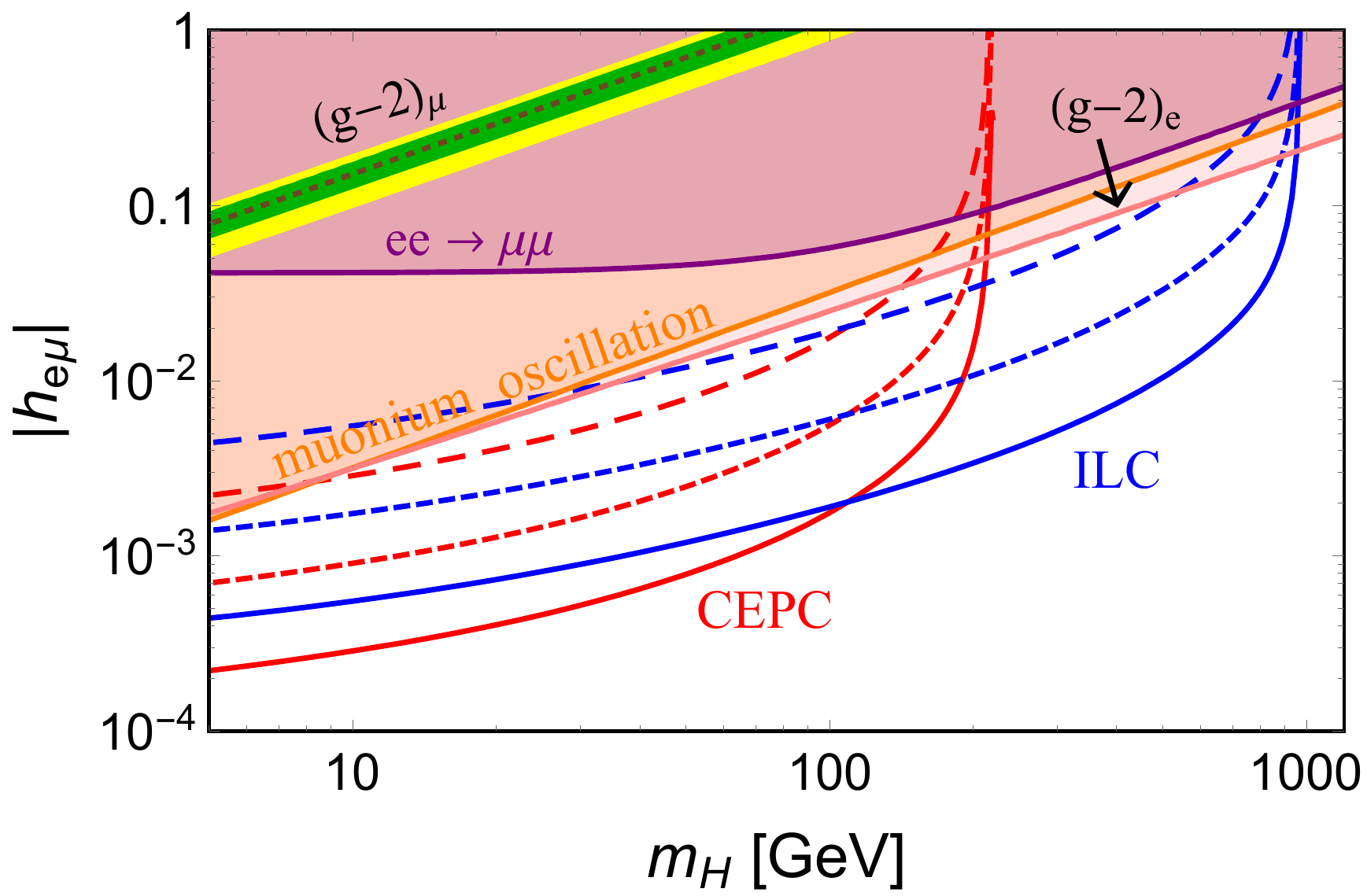}
  \includegraphics[width=0.32\textwidth]{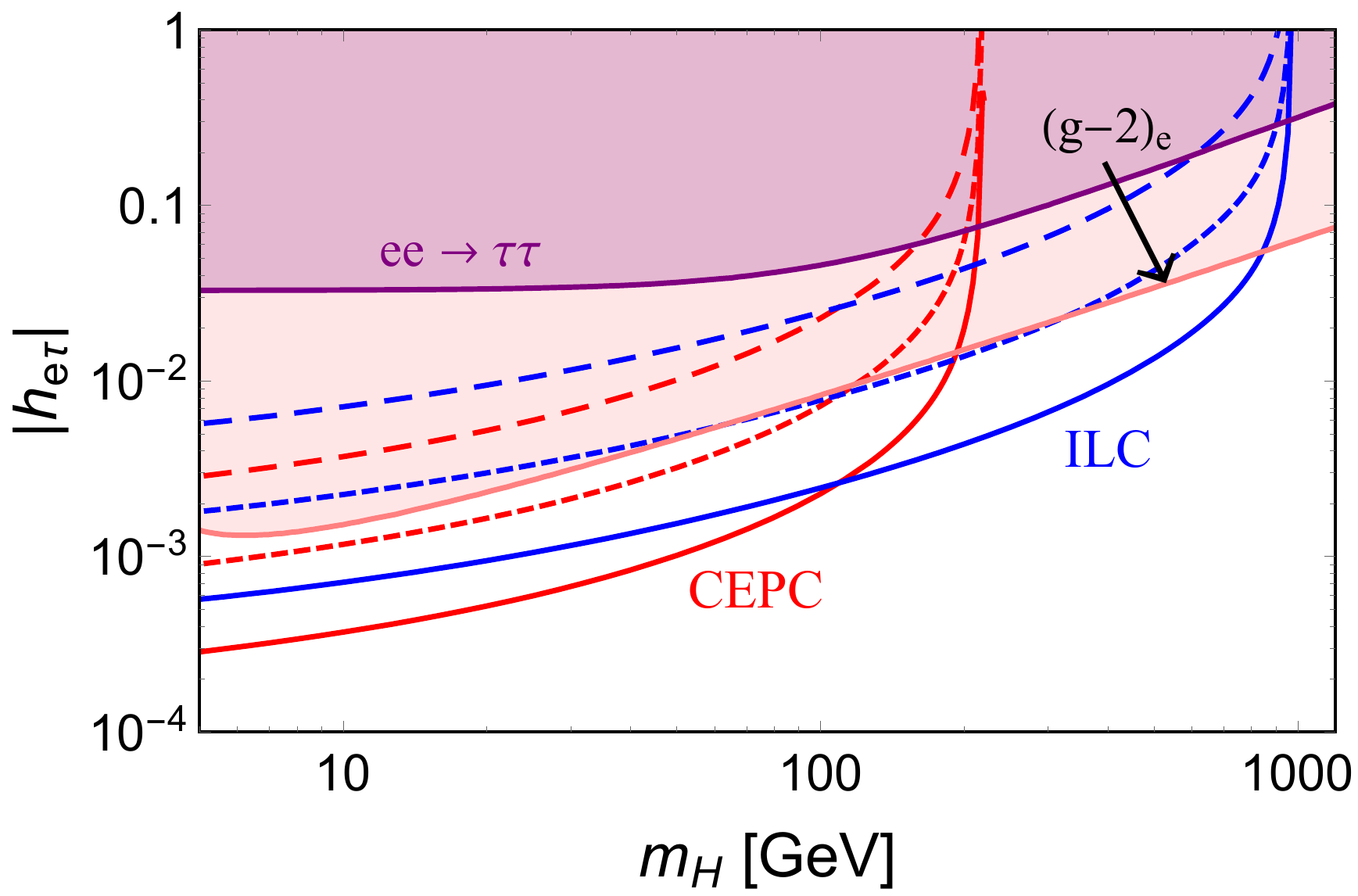}
  \includegraphics[width=0.32\textwidth]{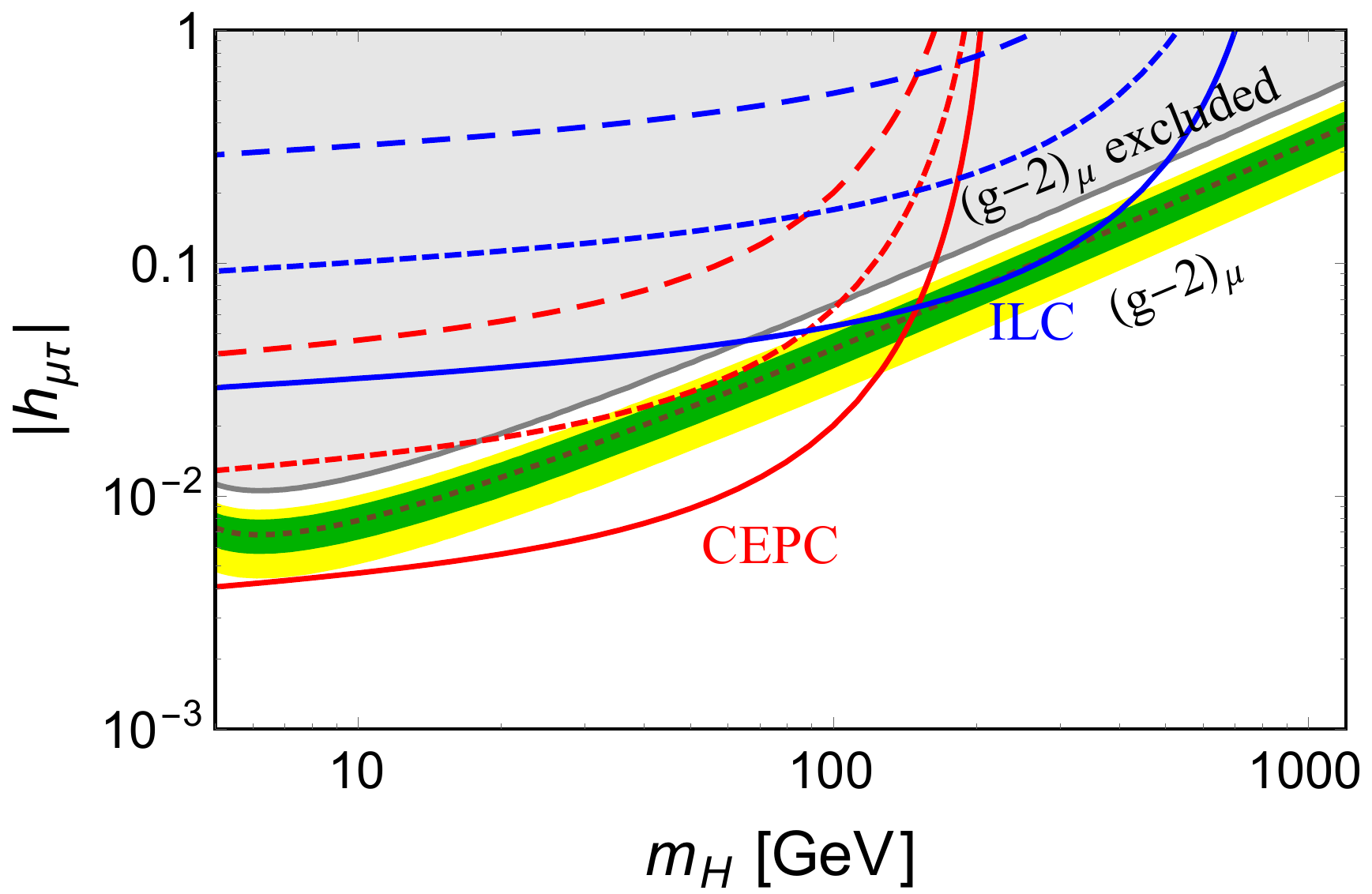}
  \vspace{-5pt}
  \caption{Prospects of LFV couplings $h_{\alpha\beta}$  from the process (\ref{eqn:H}) at CEPC 240 GeV with \mbox{5\,ab$^{-1}$} (red) and ILC 1 TeV with 1 ab$^{-1}$ (blue). A branching fraction of 1\% (long-dashed), 10\% (short-dashed) or 100\% (solid) from $H$ decay is assumed to be visible. The shaded regions are excluded by the corresponding limits. In the left and right panels, the green and yellow bands cover the $1\sigma$ and $2\sigma$ ranges of $\Delta a_\mu$. See text for more details. Figure from Ref.~\cite{Dev:2017ftk}. }
  \label{fig:H:onshell}
\end{figure}

The couplings $h_{\alpha\beta}$ of the neutral scalar $H$ can induce very rich LFV signals at high-energy colliders. In light of their clean experimental conditions, future e$^+$e$^-$ colliders are the primary facilities to search for such smoking-gun signals.
%
%
Given a single LFV coupling $h_{\alpha\beta}$ with $\alpha\neq \beta$, if kinematically allowed, the scalar $H$ can be on-shell produced at high-energy e$^+$e$^-$ colliders via the process
\begin{eqnarray}
\label{eqn:H}
\mbox{e$^+$e$^-$} \to \ell_\alpha^\pm \ell_\beta^\mp H \quad (\alpha \neq \beta) \,,
\end{eqnarray}
which is similar to the process (\ref{eqn:Zprime2}) for the Z$'$.
For simplicity, all other Yukawa couplings in the matrix $h_{\alpha\beta}$ are assumed to be vanishing. As a result, a single coupling $h_{\alpha\beta}$ cannot induce LFV decays, such as $\ell_\beta \to \ell_\alpha\gamma$ and $\ell_\beta \to 3\ell_\alpha$, which depend on the combinations of $h_{\alpha\beta}$ with other Yukawa couplings, e.g.\ $h_{\alpha\alpha}h_{\alpha\beta}$. It turns out that only a few precise LFV measurements can be used to set limits on the single coupling $h_{\alpha\beta}$. The current stringent limits on the LFV couplings $h_\mathrm{e\mu}$, $h_\mathrm{e\tau}$ and $h_{\mu\tau}$ are shown as the shaded regions in Fig.~\ref{fig:H:onshell}, including those from the anomalous magnetic moment of electron~\cite{Mohr:2015ccw} and muon~\cite{Muong-2:2006rrc, Muong-2:2021ojo}, muonium-antimuonium oscillation~\cite{Willmann:1998gd}, and the LEP e$^+$e$^- \to \ell^+ \ell^-$ data~\cite{DELPHI:2005wxt}. The prospects of the LFV couplings $h_\mathrm{e\mu}$, $h_\mathrm{e\tau}$ and $h_{\mu\tau}$ at future lepton colliders are shown respectively in the left, middle and right panels of Fig.~\ref{fig:H:onshell}, where the red and blue lines are, respectively, for the CEPC 240 GeV with an integrated luminosity of 5 ab$^{-1}$ and the ILC 1 TeV with 1 ab$^{-1}$~\cite{Dev:2017ftk}. The long-dashed, short dashed and solid lines correspond, respectively, to branching fractions of $H \to \text{e}^\pm \mu^\mp$ to be 1\%, 10\% and 100\%. It should be noted that, with the coupling $h_{\mu\tau}$, the neutral scalar $H$ provide a simple explanation for the muon $g-2$ anomaly via the $H$--$\tau$ loop. The corresponding $1\sigma$ and $2\sigma$ ranges of $\Delta a_\mu$ are presented in the right panel of Fig.~\ref{fig:H:onshell} as the green and yellow bands, respectively. It is promising that the explanation of muon $g-2$ discrepancy can be directly tested at future lepton colliders.

\begin{figure}[!t]
  \centering
  \includegraphics[width=0.32\textwidth]{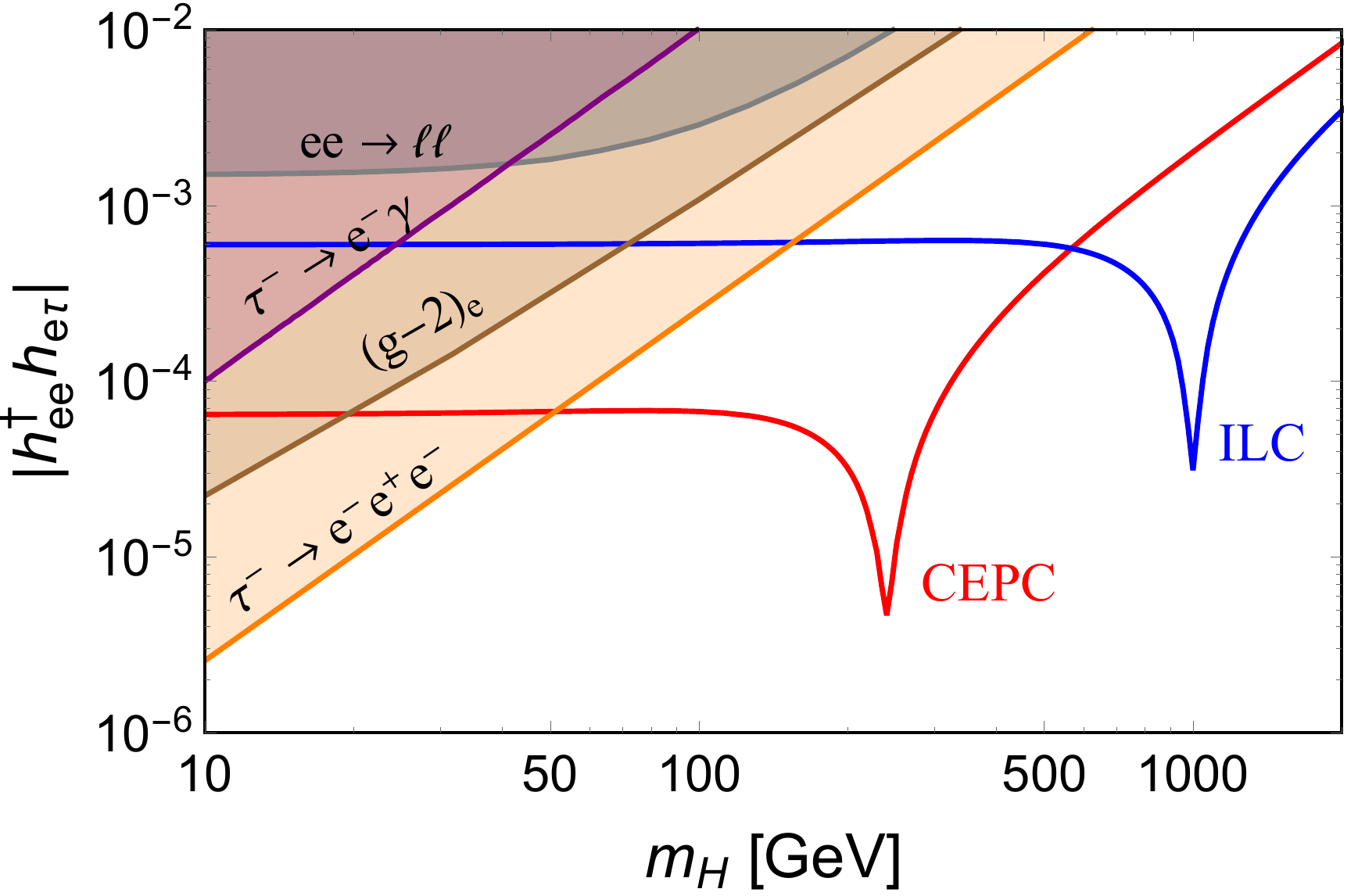}
  \includegraphics[width=0.32\textwidth]{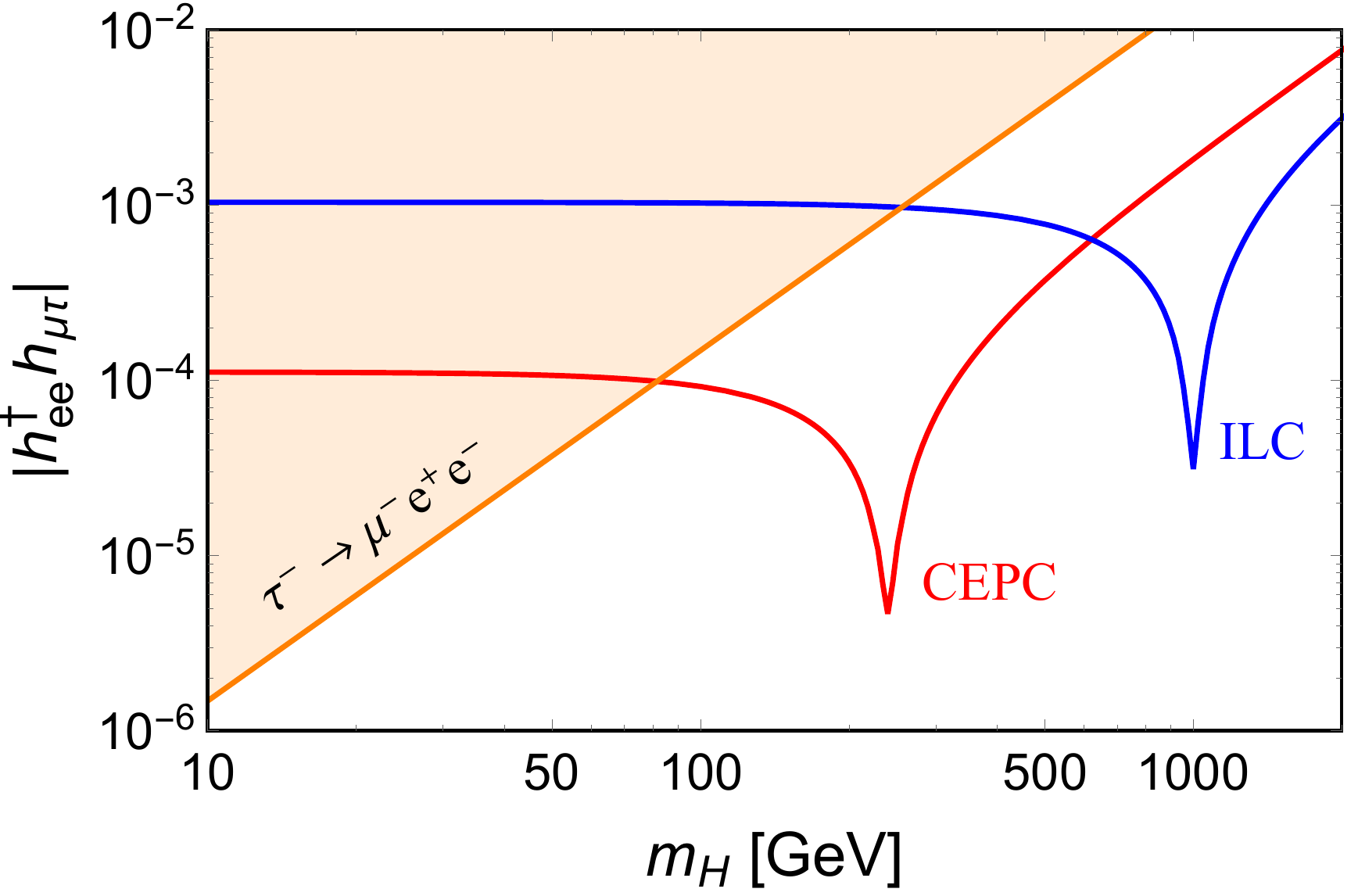}
  \includegraphics[width=0.32\textwidth]{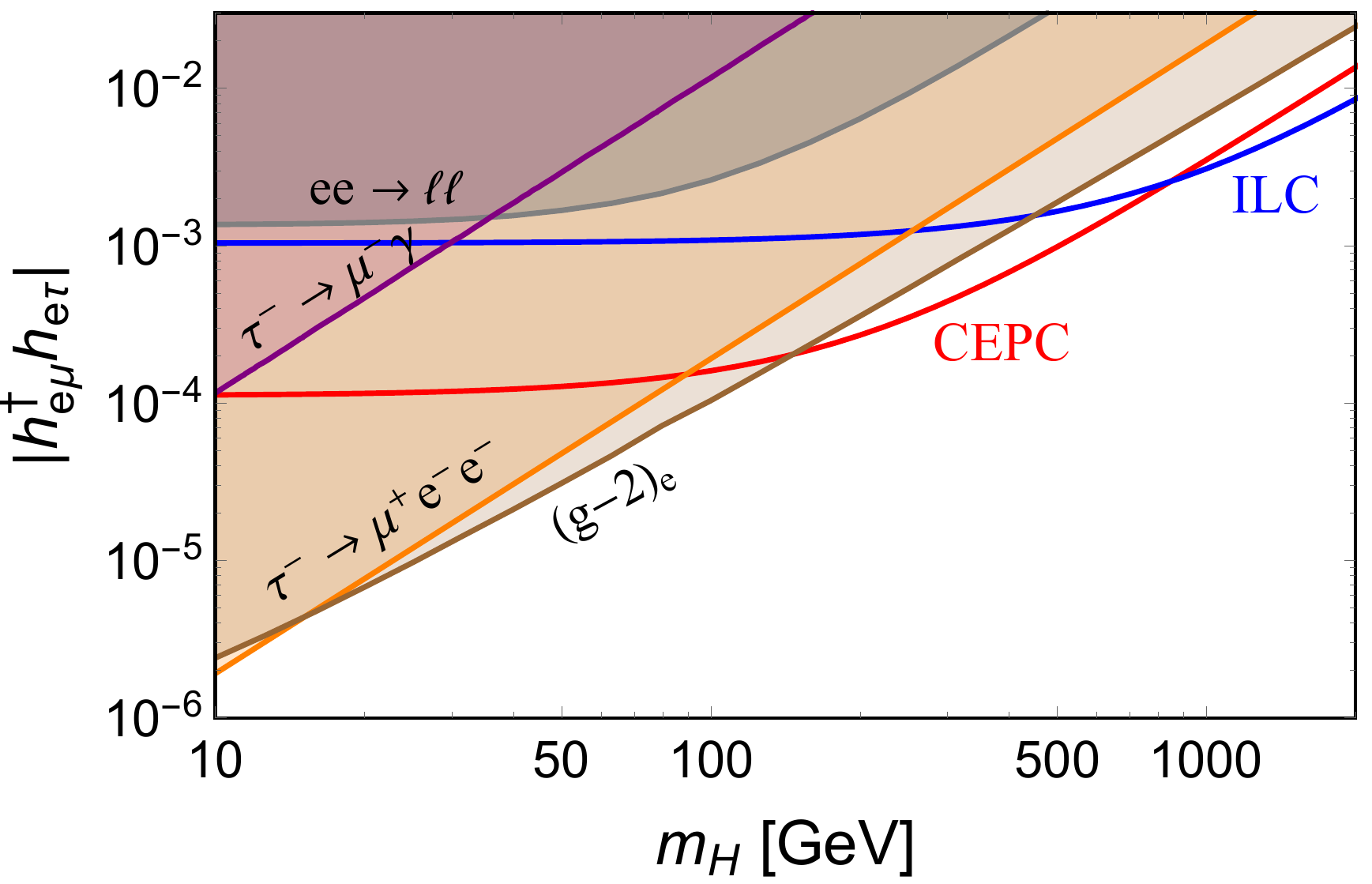}
  \vspace{-5pt}
  \caption{Prospects of $|h^\dagger_\mathrm{ee} h_{\mathrm{e}\tau}|$ (left), $|h^\dagger_\mathrm{ee} h_{\mu\tau}|$ (middle) and $|h^\dagger_{\mathrm{e}\mu} h_{\mathrm{e}\tau}|$ (right) from searches of $\mathrm{e^+ e^-} \to \mathrm{e}^\pm \tau^\mp,\, \mu^\pm \tau^\mp$ at CEPC 240 GeV with 5 ab$^{-1}$ (red) and ILC 1 TeV with 1 ab$^{-1}$ (blue). The shaded regions are excluded by the corresponding limits. See text for more details. Figure from Ref.~\cite{Dev:2017ftk}.  }
  \label{fig:H:offshell}
\end{figure}

If two Yukawa couplings are nonzero and at least one of them is LFV, for instance the combinations $h_\mathrm{ee} h_\mathrm{e\mu}$, $h_\mathrm{ee} h_\mathrm{e\tau}$, $h_\mathrm{ee} h_{\mu\tau}$ and $h_\mathrm{e\mu} h_{e\tau}$, the scalar $H$ will induce the LFV process 
\begin{eqnarray}
\mathrm{e^+ e^-} \to \ell_\alpha^\pm \ell_\beta^\mp \;\; (\alpha \neq \beta)
\end{eqnarray}
at high-energy lepton colliders. The prospects of $h_\mathrm{ee}h_\mathrm{e\mu}$ at future lepton colliders have been precluded by the stringent limit from $\mu \to \mbox{eee}$~\cite{SINDRUM:1987nra}. In the $\tau$ sector, the limits are relatively much weaker. The coupling combination $h_\mathrm{ee} h_\mathrm{e\tau}$ of $H$ can induce the process $\mathrm{e^+ e^-} \to \mathrm{e}^\pm \tau^\mp$, while the two combinations $h_\mathrm{ee} h_{\mu\tau}$ and $h_\mathrm{e\mu} h_\mathrm{e\tau}$ will both induce the process $\mathrm{e^+ e^-} \to \mu^\pm \tau^\mp$ (respectively, $s$- and $t$-channel). The prospects of these couplings at CEPC 240 GeV and ILC 1 TeV are shown in Fig.~\ref{fig:H:offshell} as the red and blue lines, respectively, requiring at least ten signal events. All the shaded regions are excluded by the limits from the rare LFV tau decays, electron $g-2$ and the LEP $\mathrm{e^+ e^-} \to \ell^+ \ell^-$ data. The dips in the left and middle panels of Fig.~\ref{fig:H:offshell} are due to the resonant production of $H$ at CEPC and ILC with $m_{H} \simeq \sqrt{s}$. 
In the limit of $m_H \gg \sqrt{s}$, the high-energy lepton colliders probe the effective four-fermion interaction $(\mathrm{\overline{e}e)}(\overline{\ell}_\alpha \ell_\beta)/\Lambda^2$ with the cut-off scale $\Lambda \simeq m_H/\sqrt{|h_{}^\dagger h_{}|}$~\cite{Kabachenko:1997aw, Cho:2016zqo, Ferreira:2006dg, Aranda:2009kz, Murakami:2014tna}.

\section{LHC and prospects for HL-LHC}
\label{sec:lhc}

During LHC Run\,2, in the period 2015--2018, the ATLAS and CMS experiments both collected nearly 140\,fb$^{-1}$ of 13\,TeV proton--proton collision data. These large data samples, corresponding to the production, per experiment, of about $8\times 10^6$ Higgs bosons, $10^8$ $t\bar{t}$ pairs, and $8\times 10^9$ Z bosons, form the basis for very sensitive searches for LFV processes with e$\mu$, e$\tau$, or $\mu\tau$ final states. With the High-Luminosity programme (HL-LHC), scheduled for 2029--2038, the integrated luminosity is expected to increase significantly reaching eventually a total of 3\,ab$^{-1}$ per experiment. 

Decays to e$\mu$ final states have a clear experimental signature, and the invariant mass is readily derived form the lepton momenta. For the e$\tau$ and $\mu\tau$ channels, the $\tau$ either decays hadronically, $\tau_h$, or leptonically, $\tau_\ell$, and the presence of either one or two neutrinos in the final state makes the reconstruction of the dilepton invariant mass challenging. Its reconstruction is based on the collinear approximation, where the direction of neutrino momentum is approximated by the direction of the other, visible decay products. The component of the event $\vec{p}_\mathrm{T}$ (transverse missing momentum) in that direction is then used as an estimate of the neutrino momentum. 
At the relevant high mass scales, where the $\tau$-leptons are typically highly boosted, the \emph{collinear mass} calculated this way is a good estimate.


\subsection{Z decays}
\label{sec:lhcZ}

\begin{figure}[tb]
 \centering
   \includegraphics[width=0.42\textwidth,
    height=0.30\textheight]{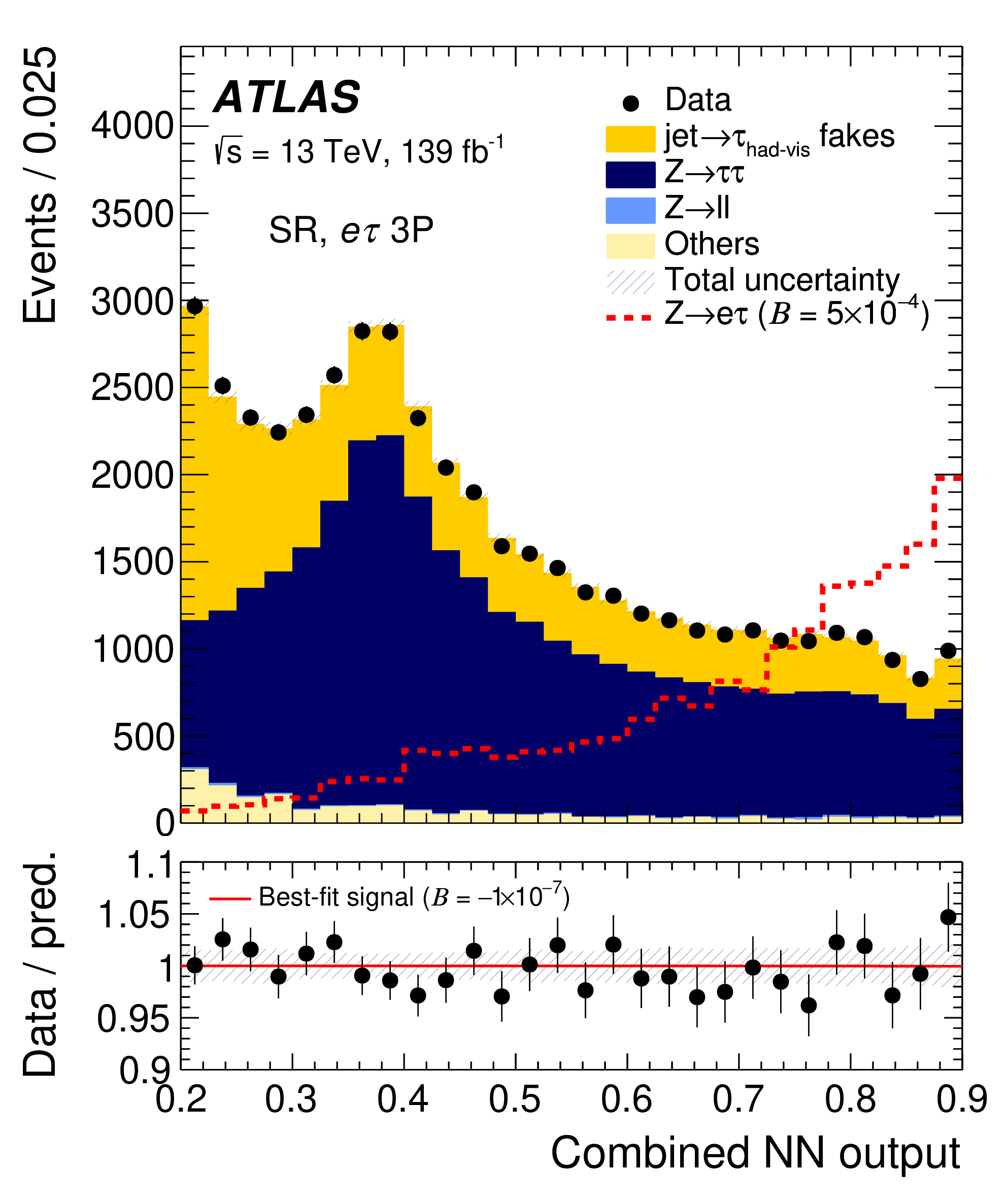}
   \quad \quad \quad
   \includegraphics[width=0.42\textwidth,   height=0.30\textheight]{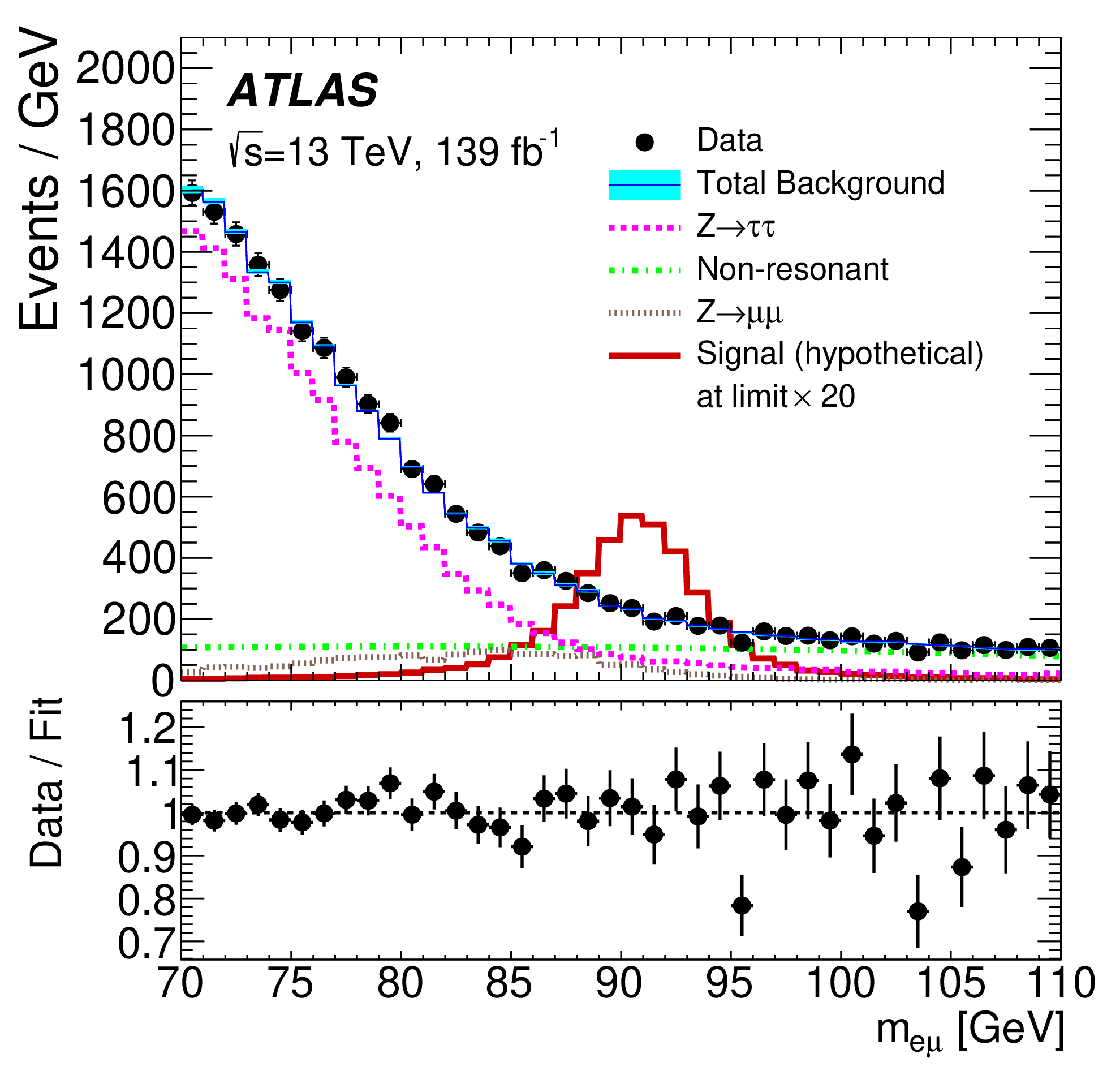} 
  \caption{ATLAS LFV Z decay search.
  Left: Observed and best-fit predicted distributions in the signal region for the $\text{Z}\rightarrow \ell \tau_\mathrm{had}$ search. Distributions of the combined NN output are shown  for the  $\text{e}\tau_\mathrm{had}$ channel, where taus decay into three charged hadrons. 
  Right: Distribution of the invariant mass of $\text{Z}\rightarrow \text{e}\mu$ candidates, for data (points) and expected backgrounds (lines) after the background-only likelihood fit. 
  }
  \label{fig:ATLAS-LFV-Z}
\end{figure}

The decay $\text{Z} \rightarrow \text{e}\mu$ has been searched for by ATLAS in their full Run-2 dataset~\cite{ATLAS:2022uhq}. The clean signature of a final state with an e$\mu$ pair whose invariant mass equals the Z mass, allowed to distinguish clearly signal from the steeply falling background of $ \text{Z} \rightarrow \tau\tau$ events, with the two $\tau$s decaying as $\tau\to\mathrm{e}\bar{\nu}\nu$ and $\tau\to\mu\bar{\nu}\nu$, and the flat combinatorial background from multijet events, where the two final state leptons were either genuine or fake. Also $\text{Z}\rightarrow \mu \mu$ events where one muon emitted bremsstrahlung, and was consequently mis-identified as an electron, was a non-negligible  background. To reduce the backgrounds, a multivariate boosted decision tree (BDT) was trained to distinguish between signal and background events, and a threshold on the BDT output was selected to optimize the
search significance. The overall search efficiency, including acceptance, of the search was about 10\%.
The final invariant mass distribution is shown in \mbox{Fig.\ \ref{fig:ATLAS-LFV-Z}}.

The measurement was dominated by the statistical uncertainty being about twice as large as the total systematic uncertainty. 
The dominant contribution to the systematic uncertainty was due to the statistical uncertainty of the simulated event samples used to form histograms of the $\text{Z}\rightarrow \tau \tau$ and $\text{Z}\rightarrow \mu \mu$ backgrounds, which were applied independently to each bin, allowing the bin content to vary. The expected upper limit including this contribution was 9.5\% higher than it would have been without this contribution.
A 95\% CL branching fraction limit was established at 
$\mathcal{B}(\mbox{Z} \to \mbox{e}\mu) < 2.62 \times 10^{-7}$.


ATLAS has published results on searches for the LFV decays $\mathrm{Z\to e\tau}$ and $\mathrm{Z \to \mu\tau}$, using their full Run-2 dataset, and considering both hadronic~\cite{ATLAS:2020zlz} and leptonic~\cite{ATLAS:2021bdj} $\tau$-lepton decay modes.
In all cases, the two final state leptons were required to be of opposite charge and, in order to reduce backgrounds from $t\bar{t}$ events, no $b$-flavoured jets were allowed. In the case of leptonic tau decays, the two  final state leptons were required to be of different flavour, \emph{i.e.}\ a e$\mu$ pair.

In the case of hadronic tau decays, the main backgrounds were the irreducible component $\text{Z}\rightarrow\tau \tau \rightarrow \ell + \mbox{\emph{hads}} + 3\nu$ and reducible backgrounds from W+jets, $t\bar{t}$ and $\text{Z}\rightarrow \ell\ell$ production, where either a jet or a light lepton ($\ell = \mathrm{e}, \mu$) was misidentified as a hadronic tau decay. In the case of leptonic tau decays, there was an additional reducible background contribution from di-bosons events.

Neural network (NN) classifiers were used to identify the three major background components: $\text{Z} \rightarrow \tau\tau$, W+jets, and $\text{Z} \rightarrow \ell\ell$, for hadronic tau decays, and $\text{Z} \rightarrow \tau\tau$, $t\bar{t}$ and di-boson production, for leptonic tau decays. 
The NN classifiers were trained and validated with the MC simulations. Components of the momenta of the light lepton and the $\tau_h$ visible momenta, their collinear mass and missing transverse momentum were typically used as the NN input variables. In order to obtain a single discriminating quantity, the combined NN classifier was built of individual NN outputs trained each against one of the three main backgrounds. An example of the combined NN classifier is shown in \mbox{Fig.\ \ref{fig:ATLAS-LFV-Z}}.

To minimise systematic uncertainties, the simulation of Z-boson production was improved with a correction derived from 
$\text{Z}\rightarrow ee,\,\mu\mu$ events. The simulated Z boson $p_\mathrm{T}$ spectra were reweighed to match the unfolded measured distribution. Furthermore, control regions, derived by inverting some of the signal-region selection cuts, were exploited in a combined fit to the NN score in the signal region, to improve uncertainties related to hadronic tau decays, and derive remaining necessary normalisation factors for the Z and top backgrounds. In the final state with hadronic tau decays, jets faking 
$\tau_h$ signatures were still significant at high NN score, where the sensitivity to LFV decays was maximal, and their normalisation was derived from the combined fit.

In the scenario where the tau leptons are unpolarized, 95\% CL branching fraction limits were established at 
$\mathcal{B}(\mbox{Z} \to \mbox{e}\tau) < 5.0 \times 10^{-6}$ and 
$\mathcal{B}(\mbox{Z} \to \mu\tau) < 6.5 \times 10^{-6}$.
For the establishment of the search sensitivity, the statistical uncertainty contribution was found to be dominant. The systematic uncertainties on the tau lepton energy and identification in the hadronic tau final state and on the transverse missing energy in the leptonic tau final state are unlikely to be further reducible.


Since the ATLAS LFV Z searches have backgrounds, the limits can be expected to improve by a factor $1/\sqrt{\mathcal{L}}$, where $\mathcal{L}$ is the collected luminosity, implying ultimately a factor of five improvement down to sensitivities to branching fractions somewhat \mbox{below $10^{-6}$} for $\text{Z}\rightarrow \ell\tau$ and \mbox{below $10^{-7}$} for $\text{Z}\rightarrow \text{e}\mu$. The increase in the ATLAS tracking detector's acceptance in the HL-LHC phase, as well as a possible combination with a similar analysis from the CMS experiment, could potentially improve the sensitivities further.


\subsection{Z$\boldsymbol{'}$ and other resonances searches}
\label{sec:lhcZprime}

Searches for heavy resonant (and non-resonant) production of e$\mu$, e$\tau$, and $\mu\tau$ final states in 13\,TeV LHC data have been reported by CMS from their full Run\,2 data set (137.1\,fb$^{-1}$)~\cite{CMS:2022fsw} and by ATLAS from a partial dataset (36.1\,fb$^{-1}$)~\cite{ATLAS:2018mrn}. No evidence for physics beyond the standard model in the dilepton invariant mass spectra was found in neither of the analyses, allowing limits to be set on various BSM models, including models with heavy Z$'$ gauge bosons with LVF transitions. 

Processes leading to LFV final states have a clear experimental signature and a low background from SM processes. Unlike for flavour-diagonal decays, there is no irreducible background from the Drell--Yan process (dilepton production in hadron--hadron collisions), and the dominant backgrounds are from $\mathrm{t\bar{t}}$ and diboson (WW, WZ, ZZ) production, and from multijet and W+jets processes, where jets are misidentified as leptons. 

For both analysis it was assumed that the Z$'$ boson has couplings to quarks similar to those of the SM Z boson, and that it decays also to LFV e$\mu$, e$\tau$, and $\mu\tau$ final states. Whereas CMS assumed the flavour-diagonal lepton couplings to be SM-like, but added a 10\% branching fraction to each of the LFV modes, ATLAS assumed that only LFV leptonic decays were allowed, and that only one of these was active at the time with a couplings strength similar to the leptonic couplings of the SM Z boson. From their larger data sample, CMS was able to set the most stringent limits, excluding a Z$'$ up to 5.0\,TeV in the e$\mu$ channel, 4.3\,TeV in the e$\tau$ channel, and 4.1\,TeV in the $\mu\tau$ channel.


\subsection{Higgs decays}
\label{sec:lhcHiggs}

Searches for LFV decays of the Higgs boson have been performed by ATLAS, CMS, and LHCb. While some of these analyses focus on the 125-GeV particle, other explore more generally LFV decays of a lighter or heavier scalar particle. 

Searches for the decay of the 125-GeV Higgs boson to e$\tau$ or $\mu\tau$
can be performed in different final states depending on the $\tau$ decay mode. Final states with hadronically decaying taus, denoted $\tau_h$, benefit from a larger branching fraction but suffer from larger backgrounds from quark- and gluon-jets.  

ATLAS and CMS have both explored the e$\mu$, e$\tau_h$, and $\mu\tau_h$  final states~\cite{ATLAS:2019pmk,CMS:2021rsq}. Events were selected online with trigger paths based on the presence of an electron or a muon with sufficient transverse momentum, typically 30\,GeV. 
The dominant backgrounds consisted of $\mbox{Z} \to \tau\tau$ events,
$\mbox{W} (\to \ell\nu) + \mbox{jets}$ events
with a jet misidentified as a lepton, QCD multijet events with two misidentified jets, and $t\bar{t}$ and diboson productions. To increase the sensitivity, events were divided into several categories with different signal-to-background ratios. While the choice of the analysis categories depended on the experiment, both collaborations isolated events with two jets with a large invariant mass, characteristic of the vector-boson-fusion production mode. 

\begin{figure}[tb]
 \centering
  \includegraphics[width=0.49\textwidth]{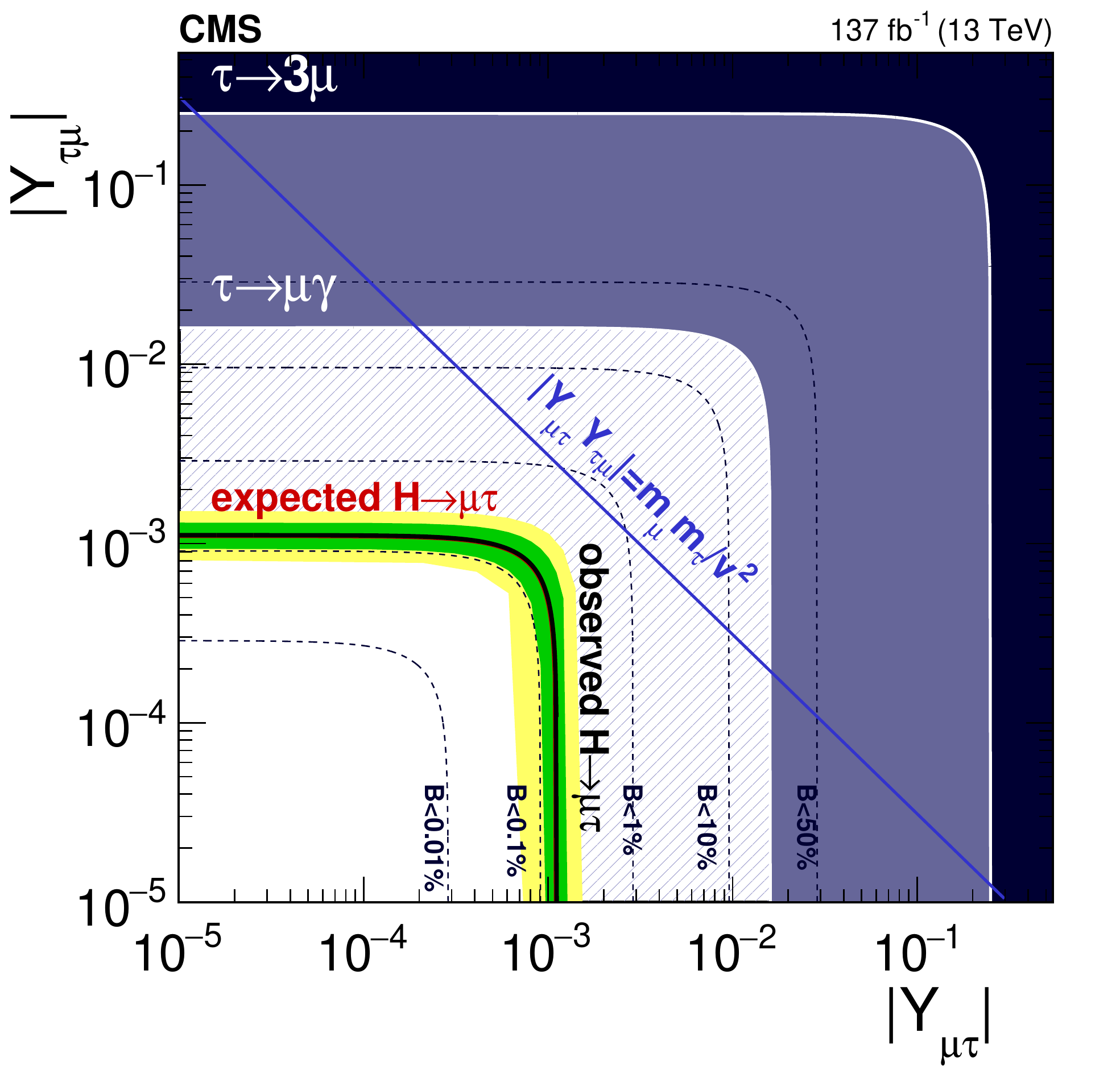}
  \includegraphics[width=0.49\textwidth]{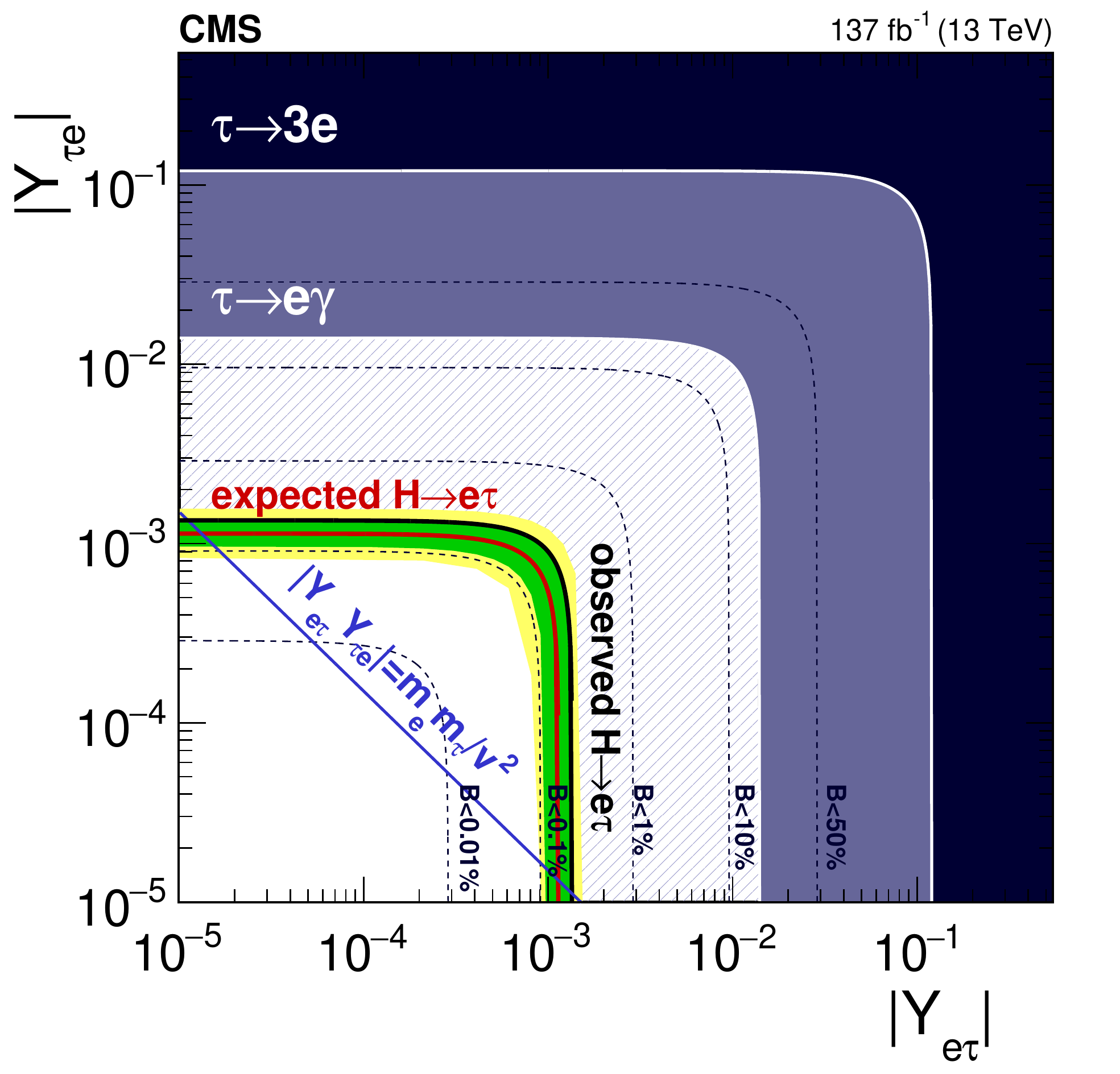}
  \caption{CMS $H \to \ell\tau$ search: Expected (red line) and observed (black solid line) 95\% CL upper limits on $|Y_{\mu\tau}|$ vs.\ $|Y_{\tau\mu}|$.  The green and yellow bands indicate the range containing 68\% and 95\% of all observed limit variations from the expected limit. The shaded regions are constraints obtained from null searches for $\tau\to3\mu$ or $\tau\to 3e$ (dark blue) and $\tau\to\mu\gamma$ or $\tau\to e\gamma$ (purple). The blue diagonal line is the theoretical naturalness limit $|Y_{ij}Y_{ji}|=m_im_j/v^2$  (from Ref.\cite{CMS:2021rsq}).}
  \label{fig:Ycms}
\end{figure}

In a CMS analysis, based on the full Run-2 dataset, 
results were extracted from a maximum likelihood fit of the distribution of a boosted decision tree (BDT) output in the different categories of each final state. The BDT was trained against a mixture of the dominant backgrounds and included variables such as the transverse momentum of the leptons, the collinear mass, the visible mass of the dilepton system, the transverse mass between the missing transverse momentum and either of the leptons, the pseudorapidity difference between the leptons, the azimuthal-angle difference between the leptons, and the azimuthal-angle difference between the missing transverse momentum and either of the leptons. No excess over the expected background was oberved, and 95\% CL limits on the Yukawa couplings were established at $\sqrt{|Y_{\mu\tau}|^2 + |Y_{\tau\mu}|^2} <1.11\times 10^{-3}$ and $\sqrt{|Y_{e\tau}|^2 + |Y_{\tau e}|^2} <1.35\times10^{-3}$, as shown in \mbox{Fig.\ \ref{fig:Ycms}}. The limits on the branching fractions were $\mathcal{B}(H\to \mu\tau)<0.15\%$ and $\mathcal{B}(H\to e\tau)<0.22\%$. 

ATLAS has currently analyzed 36 fb$^{-1}$ of 13\,TeV data collected in 2016~\cite{ATLAS:2019pmk}, and has observed no excess over the background expectation either.
A BDT was used to separate the signal from the backgrounds in the different signal regions. The collinear mass as well as the dilepton mass reconstructed with the MMC algorithm were used as input variables, among other variables. The backgrounds from top-quark and Z-boson productions were constrained from dedicated control regions in data. The 95\% CL branching fraction limits were $\mathcal{B}(H\to\mu\tau)< 0.28\%$ and $\mathcal{B}(H\to \mbox{e}\tau) < 0.47\%$.
 
Significantly more stringent limits are expected to be set at the HL-LHC, thanks to the increased integrated luminosity, as well as improvements to the object reconstruction (in particular for hadronically decaying taus) and analysis strategy (e.g.\ using deep neural network or using more production information). Theory uncertainties for the signal are also expected to be reduced \cite{Cepeda:2019klc}. Taking these points into account, a limit in the range 0.01--0.05\% could potentially be reached on the $\mathcal{B}(H\to\mu\tau)$ branching fraction.

\begin{figure}[tb]
 \centering
  \includegraphics[width=0.6\textwidth]{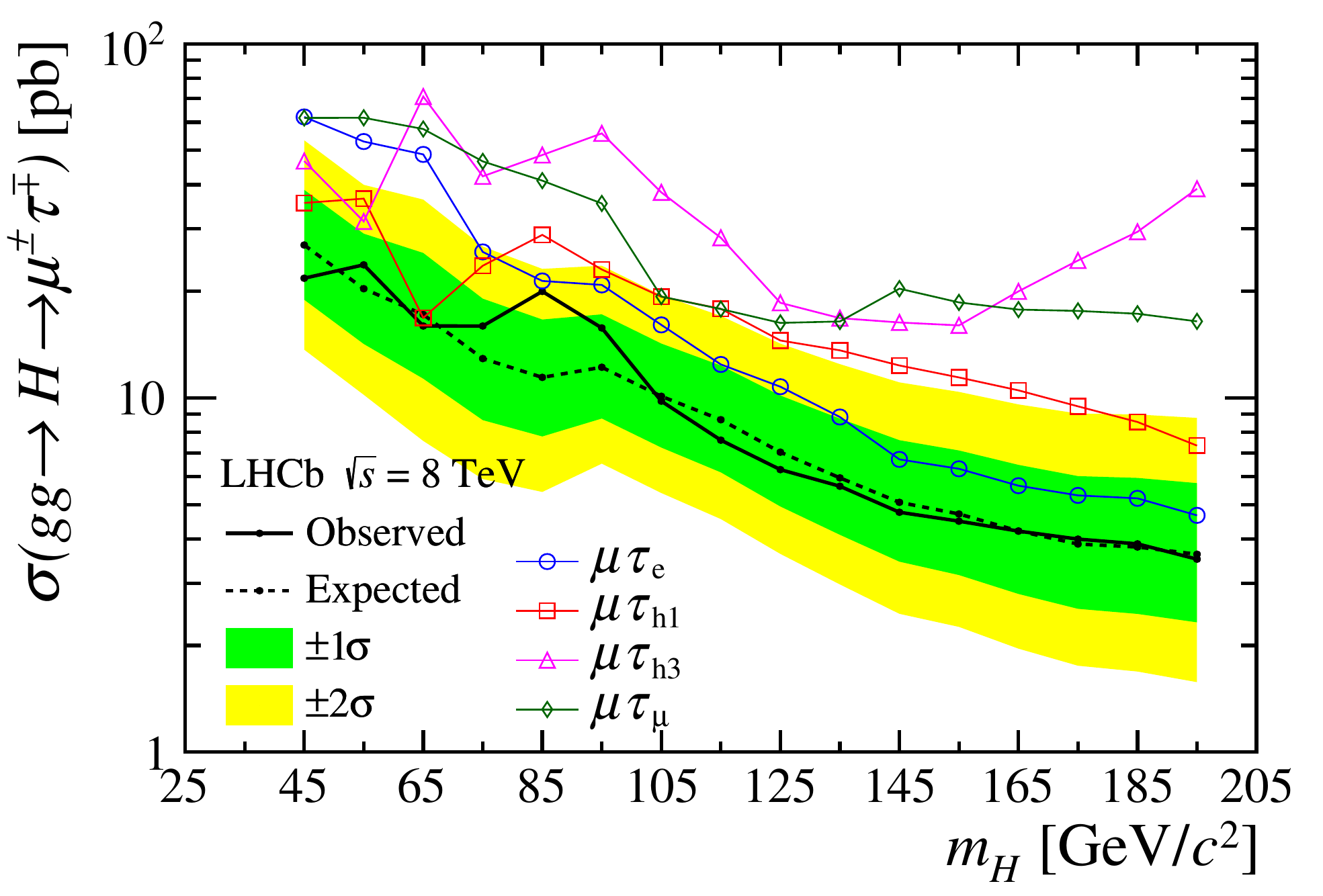}
  \caption{LHCb search for LFV scalar particle decays, $H\to\mu\tau$:
Cross-section times branching fraction 95\% CL limits as a function of $m_H$. The observed limits from individual channels are also shown (from Ref.\cite{LHCb:2018ukt}).}
  \label{fig:lhcbmutau}
\end{figure}

LHCb has also explored the possibility of LFV decays of a scalar particle to $\mu\tau$ final states, scanning over a large mass range of 45--\mbox{195\,GeV}, using 2\,fb$^{-1}$ of 8\,TeV data~\cite{LHCb:2018ukt}. The analysis targets scalar particles produced in the forward region via the gluon-gluon fusion process. No excess over the background estimate was observed.
An upper limit on the production cross-section multiplied by the branching fraction was set and ranges from 22\,pb for a boson mass of 45\,GeV to 4\,pb for a mass of 195\,GeV at 95\% CL, as shown in \mbox{Fig.\ \ref{fig:lhcbmutau}}. While the branching fraction limit at a mass of 125\,GeV ($\mathcal{B}(H\to\mu\tau)<26\%$) is weaker than those of ATLAS and CMS, LHCb provides the most sensitive results so far for scalar masses below 125\,GeV.

CMS has scanned the higher mass spectrum from 200 to 900\,GeV in a search for decays of a heavy scalar to e$\tau$ or $\mu\tau$ final states, using 36 fb$^{-1}$ of 13\,TeV data~\cite{CMS:2019pex}. The analysis strategy was close to that focussed on the 125-GeV particle, except that more emphasis was put on the precise reconstruction of the dilepton mass spectrum,
and fewer model-dependent assumptions on the production of the scalar particle were made. The upper limits on the production cross section multiplied by the branching fraction range from 51.9 to 1.6\,fb for the $\mu\tau$ mode and from 94.1 to 2.3\,fb for the $e\tau$ mode, at 95\% CL.

\begin{figure}[tb]
 \centering
  \includegraphics[width=0.45\textwidth]{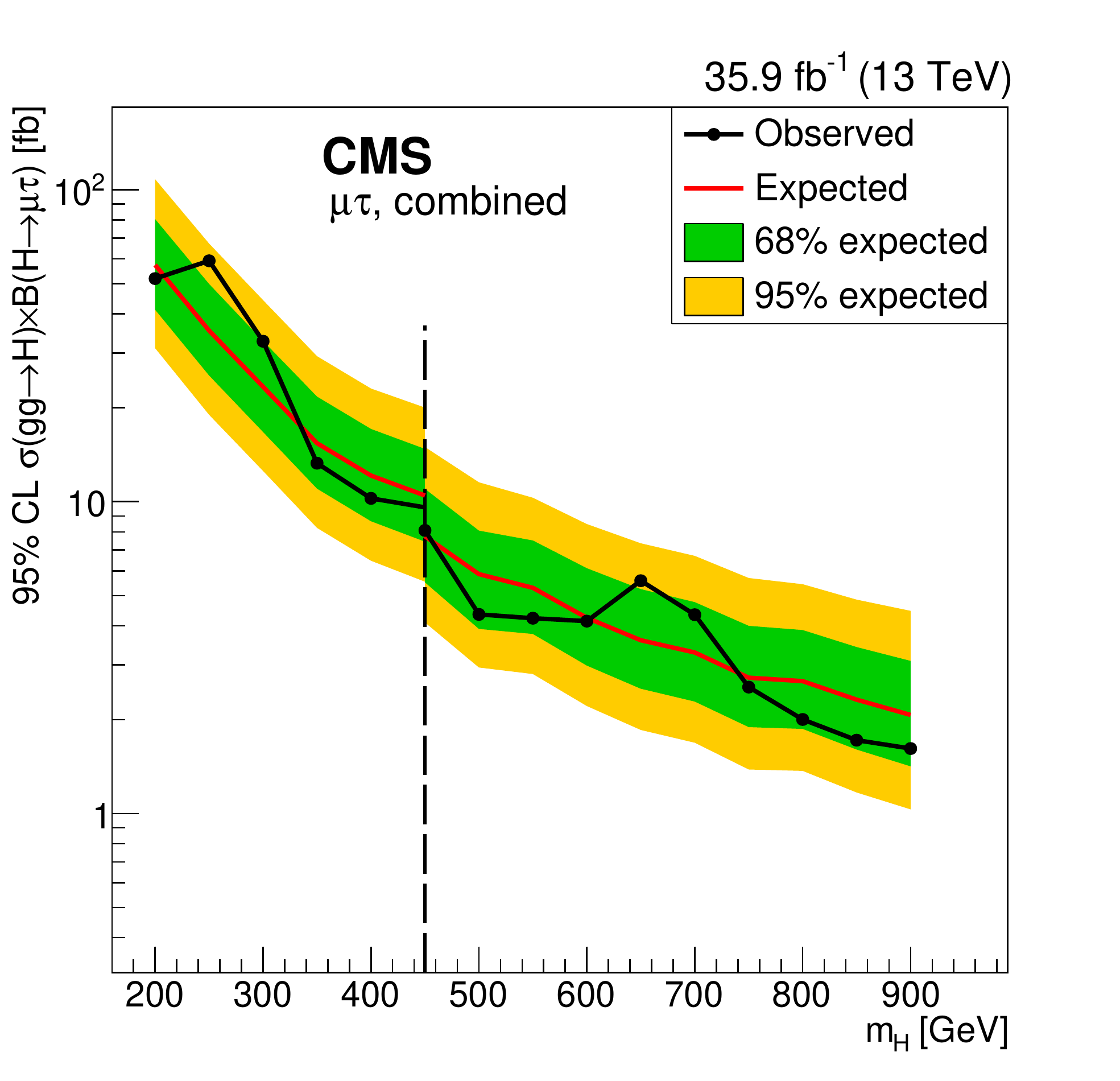}
  \includegraphics[width=0.45\textwidth]{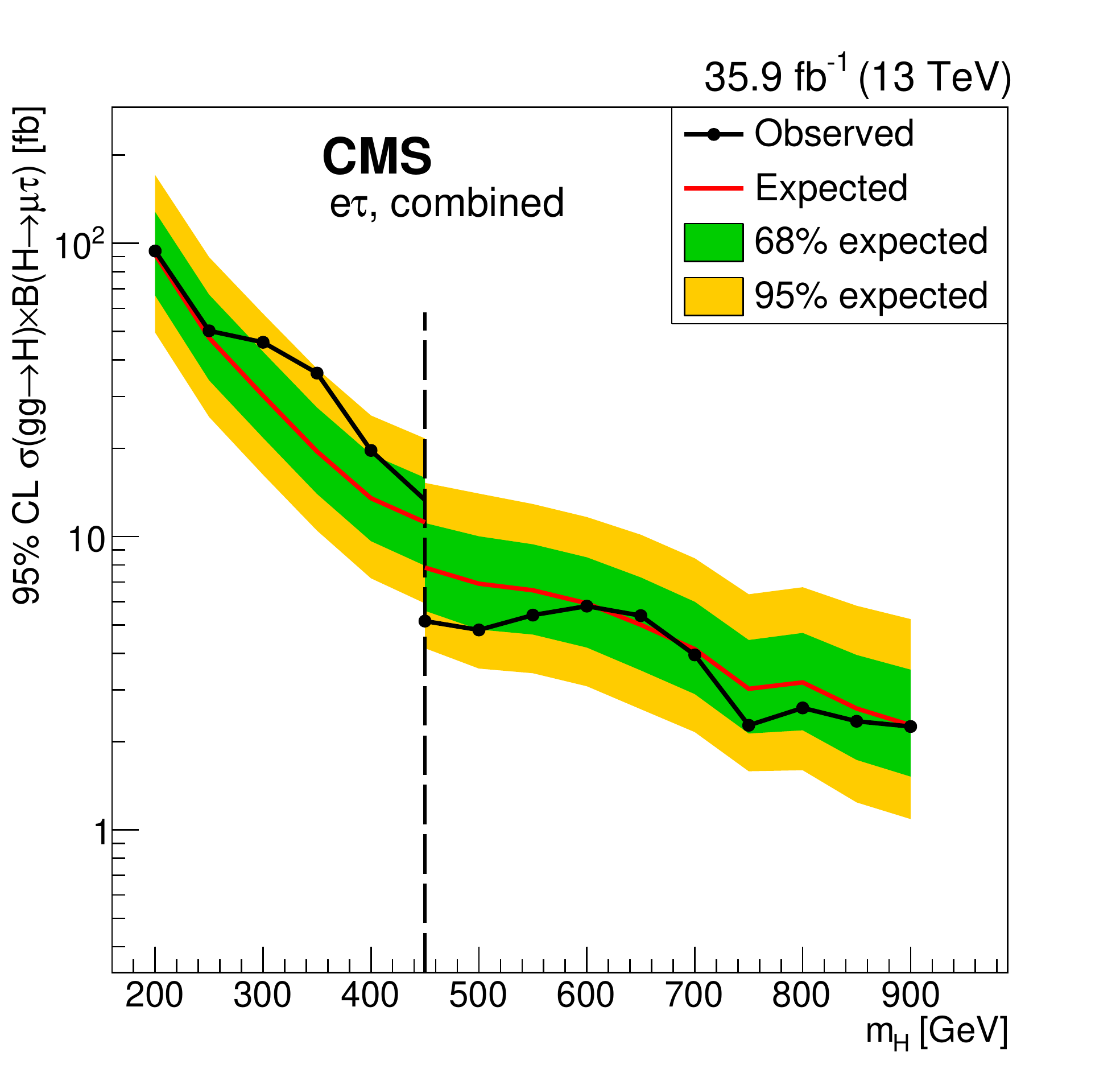}
  \caption{CMS search for LFV decays of a heavy scalar particle, $H\to\ell\tau$: Observed and expected 95\% CL upper limits on the production cross section multiplied by the branching fraction for the $\mu\tau$ (left) and $e\tau$ (right) decay channels
  (From Ref.\cite{CMS:2019pex}).}
  \label{fig:cmshig18017}
\end{figure}

Searches for Higgs boson decays to e$\mu$ can take advantage of the excellent mass resolution of the lepton pair, and look for a narrow excess of events
over a smooth background. The ATLAS experiment has established the most stringent limit on such a decay, $\mathcal{B}(H\to e\mu)<6.1\times10^{-5}$,
from their full Run-2 dataset \cite{ATLAS:2019old}. In the analysis, events were classified in various categories to improve the analysis sensitivity, on the basis of the following variables: the subleading lepton transverse momentum, the dilepton transverse momentum, the lepton pseudorapidity, and the dijet pseudorapidity and invariant mass if there were at least two jets. The results were extracted from an unbinned parametric maximum likelihood fit, where the narrow signal resonance was parameterized by the sum of a Crystal Ball function and a Gaussian function, and a Bernstein polynomial of degree 2 was used to describe the background mass spectrum. The result is significantly better than the limit of $3.5\times 10^{-4}$ set by the CMS Collaboration using their Run-1 \mbox{dataset \cite{CMS:2016cvq}}.
\begin{figure}[tb]
 \centering
  \includegraphics[width=0.45\textwidth]{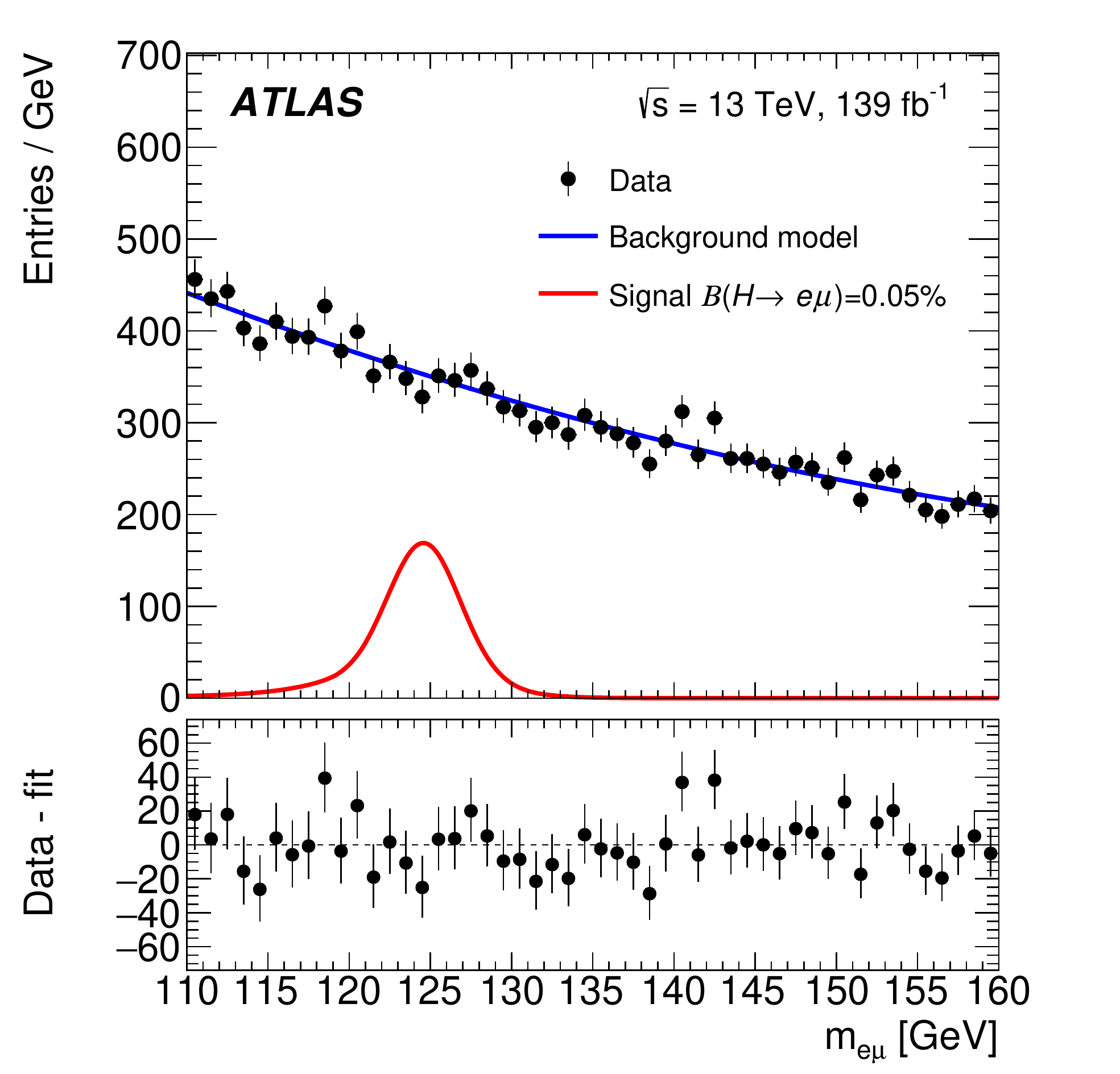}
  \caption{Dilepton invariant mass  
  for the $e\mu$ channel 
  compared with the background-only model. The signal expectation with $\mathcal{B}(H \to e\mu) = 0.05\%$ is shown in red. The bottom panel shows the difference between data and the background-only fit (from Ref.\cite{ATLAS:2019old}).}
  \label{fig:atlasMDiLept}
\end{figure}


\subsection{Top quark processes}
\label{sec:lhcTop}

A search for CLFV processes in top-quark reactions has been recently published by CMS~\cite{CMS:2022ztx}. The analysis, based on 138\,fb$^{-1}$ of 13\,TeV LHC proton-proton collision data, includes both production and decay modes of the top quark, \emph{i.e.}\ 
$\mbox{q} \to \mbox{e}\mu\mbox{t}$ 
and 
$\mbox{t} \to \mbox{e}\mu\mbox{q}$, 
respectively, with q referring to an up or charm quark. The result of the search was interpreted in terms of limits on vector, scalar, and tensor four-fermion interactions.

\begin{figure}[tb]
 \centering
  \includegraphics[width=0.5\textwidth]{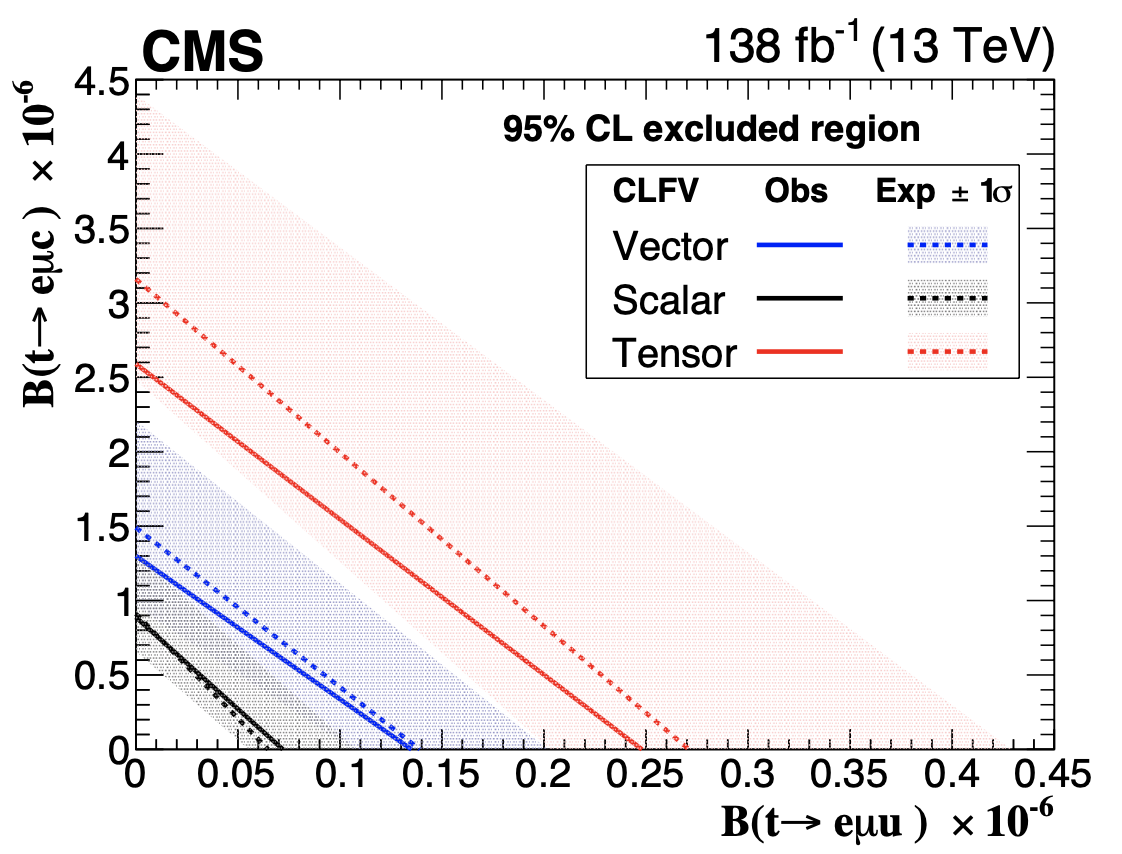}
  \caption{CMS CLFV top decay search: Observed 95\% CL exclusion limits on $\mathcal{B}(\mbox{t} \to \mbox{e}\mu\mbox{c})$ as a function of $\mathcal{B}(\mbox{t} \to \mbox{e}\mu\mbox{u})$ for the vector-, scalar-, and tensor-like CLFV interactions. The hatched bands indicate the regions containing 68\% of the distribution of limits expected under the background-only hypothesis (from Ref.~\cite{CMS:2022ztx}).}
  \label{fig:CLFV_top}
\end{figure}

Whereas the production-mode process involves single-top events, the decay-mode process is dominated by t$\bar{\mbox{t}}$ events.
In both cases, signal events contain an oppositely charged e$\mu$ pair together with multiple jets, one of which stems from the hadronization of a bottom quark from a $\mbox{t} \to \mbox{bW}$ decay. 
An event sample was selected by requiring
one electron, one muon of opposite charge, and at least one b-tagged jet.
The dominant source of background was 
from t$\bar{\mbox{t}}$ events. Hence, the sample was divided into a signal region containing events with exactly one b-tagged jet and a t$\bar{\mbox{t}}$ control region containing events with at least two b-tagged jets. 
A boosted decision tree (BDT) combining five discriminating variables was used to distinguish the signal from the background. 
The distributions of the input variables, of which the $p_\mathrm{T}$ of the leading lepton was the most powerful, were found to be well described by the simulated background model. 
The BDT was trained on simulated samples of CLFV signal events, on the one hand,
against background events, on the other.
Events from the 
production channel
($\mbox{q} \to \mbox{e}\mu\mbox{t}$)
were found to 
dominate over those from the 
decay channel
($\mbox{t} \to \mbox{e}\mu\mbox{q}$)
in the discrimination. Two factors contributed to this observation: Firstly, the production channel had a higher event yield than the decay channel. Secondly, the production channel was better separated from the background via a harder final-state-particle $p_\mathrm{T}$ distribution. 
%
The data was found to be consistent to expectation of the SM in the absence of signal, and upper limits were set on the strength of individual vector-, scalar-, and tensor-like four-fermion effective operators. These were then converted into 95\% CL limits on branching fractions of the top quark 
$\mathcal{B}(\mbox{t} \to \mbox{e}\mu\mbox{q})$, q=u (c),
$<0.13 \times 10^{-6}$ ($1.31 \times 10^{-6}$),
$0.07 \times 10^{-6}$ ($0.89 \times 10^{-6}$),
and $0.25 \times 10^{-6}$ ($2.59 \times 10^{-6}$)
for vector, scalar, and tensor CLFV interactions, respectively. The search limits on
$\mathcal{B}(\mbox{t} \to \mbox{e}\mu\mbox{c})$ and $\mathcal{B}(\mbox{t} \to \mbox{e}\mu\mbox{u})$ are indeed correlated, as illustrated in \mbox{Fig.\ \ref{fig:CLFV_top}}.

It is interesting to observe that the CMS limits are about one order of magnitude lower than that predicted in \mbox{Ref.\ \cite{Davidson:2015zza}} from a similarly sized data sample. 
The main reason for this difference shall be probably be traced to CMS's use of the production channel, which was not considered in the earlier work.

\section{Future \texorpdfstring{e$^+$e$^-$}{ee} colliders}
\label{sec:ee}

Electron-positron colliders are largely complementary to proton-proton colliers in the search for rare phenomena in heavy particle decays. Whereas proton-proton colliders provide enormous event samples, electron-positron colliders generally deliver smaller samples but in cleaner experimental conditions. A further advantage arises from the constrained kinematics in e$^+$e$^-$ collisions, where the initial state is precisely known. As an illustration of the complementarity between the two types of colliders, the recent ATLAS limits on CLFV Z decays, were derived from event samples with two-to-three orders of magnitude more Z decays than the previous limits from LEP, which they succeeded by small factors. 

Several future electron-positron colliders have been proposed. Until ten years ago, it was generally believed that any future high-energy e$^+$e$^-$ collider would be linear, and there were two proposals on the market: The International Linear Collider (ILC)~\cite{Bambade:2019fyw, LCCPhysicsWorkingGroup:2019fvj} and the Compact Linear Collider (CLIC)~\cite{CLICdp:2018cto}. With the discovery in 2012 
that the Higgs boson is light, circular colliders had a strong revival with two proposed projects: The FCC-ee~\cite{FCC:2018byv} at CERN and the largely similar CEPC~\cite{CEPCStudyGroup:2018ghi} in China. For the search for rare decays of heavy SM particles, circular colliders have a marked advantage in their superior luminosities at the relevant collisions energies: the Z pole ($\sqrt{s}=91.2$\,GeV), the cross-section maximum for HZ production
($\sqrt{s}=240$\,GeV), and the threshold for $\mathrm{t\bar{t}}$ production ($\sqrt{s}\gtrsim 350$\,GeV). As an illustration, where a GigaZ phase is discussed for the ILC as a possible upgrade option beyond the 15-years Higgs programme, 
a TeraZ phase is already included in the FCC-ee baseline programme.

The FCC-ee is the first stage of the integrated Future Circular Colliders (FCC) project to be based on a novel research infrastructure hosted in a $\sim$100-km tunnel in the neighbourhood of CERN. 
The FCC-ee programme~\cite{Blondel:2021ema} includes four major phases with precision measurements of the four heaviest particles of the Standard Model: 
\emph{i})~the Z boson, with $5\times 10^{12}$ Z decays collected around the Z pole \mbox{(4 years)}, 
\emph{ii})~the W boson, with $10^8$ WW pairs collected close to threshold \mbox{(2 years)}, 
\mbox{\emph{iii})}~the Higgs boson, with $1.2\times 10^6$ \mbox{e$^+$e$^- \to$ HZ} events produced at the  cross-section maximum \mbox{(3 years)}, and
\emph{iv})~the top quark, with $10^6$ $\mathrm{t\bar{t}}$ pairs produced at and slightly above threshold \mbox{(5 years)}. It is interesting to note that,
with its extreme Z-pole luminosity, exceeding $10^{36}\,\text{cm}^{-2}\,\text{s}^{-1}$, FCC-ee will produce a Z sample larger by about two orders of magnitude than that of HL-LHC, and even comparable to what is expected from FCC-hh.

FCC-ee will feature two (possible four) interaction points each equipped with a powerful, state-of-the art detector system. Detector concepts being studied feature a solenoidal magnetic field, a small pitch, thin layers vertex detector
providing an excellent impact parameter resolution for lifetime measurements, a highly transparent tracking system providing a superior momentum resolution,
a finely segmented calorimeter system with excellent energy resolution, 
and a highly efficient muon system.
At least one of the detector systems will be equipped with efficient particle-identification capabilities allowing $\pi/\text{K}$ separation over a wide momentum range. Important for the following discussion is a typical momentum resolution of $\sigma_p/p \simeq 1.5 \times 10^{-3}$ at $p=45.6$\,GeV.

\subsection{Z Decays}

The prospects of searches for CLFV Z decays at the FCC-ee were first discussed in 
\mbox{Ref.\ \cite{Dam:2018rfz}}, from which much of the following information has been extracted.

\subsubsection*{\texorpdfstring{$\boldsymbol{\text{Z} \rightarrow \mu\mathrm{e}}$}{Ztomue}}

From OPAL at LEP, a background-free search for $\text{Z} \to \mu$e decays resulted in a limit of $1.7\times 10^{-6}$ (95\% CL). Even with a much improved detector performance, backgrounds will eventually show up in the 
$\mathcal{O}(10^5)$ larger FCC-ee samples. 
As at the LHC, an irreducible background is expected from 
$\text{Z} \to \tau\tau \to \mu \text{e} \bar{\nu}\nu\bar{\nu}\nu$,
where the two charged leptons are both emitted close to the end-point of their momentum spectra. 
With the superior momentum resolution of FCC-ee detectors, 
this background
is tiny corresponding to a Z-boson branching fraction of only $\mathcal{O}(10^{-11})$.
Here, FCC-ee has an enormous advantage over LHC. Not only is the momentum resolution better, 
but, more importantly, the Z-mass constraint is immensely more powerful. Whereas in proton-proton collisions, the mass
constraint is limited by the 2.5~GeV natural width of the Z boson, at FCC-ee
the mass is precisely known from the collision energy with a dispersion of only
85~MeV from the beam energy spread.

A potentially more serious background arises from so-called catastrophic
brems\-strahlung of muons in the material of the electromagnetic
calorimeter (ECAL), by which a muon radiates off a significant fraction of its energy, with the subsequent risk of being mis-identified as an electron.
In \mbox{Ref.\ \cite{Dam:2018rfz}} it is argued that this background can be  controlled down to a level corresponding to a 
Z-boson branching fraction 
of $\mathcal{O}(10^{-8})$ via a careful design of the ECAL including a good energy resolution and longitudinal segmentation, where the latter allows 
for the suppression of late starting showers.
Important for controlling this background is the ability to precisely determine the mis-identification rate from the data themselves. For this, an independent means of e/$\mu$ separation, such at that provided by a powerful d$E$/d$x$ measurement, will be essential, and it is encouraging to notice, that a 3--4 standard deviation e/$\mu$ separation at $p = 45.6$\,GeV may be possible via the use of the cluster counting technique in a Helium based drift chamber~\cite{FGrancagnolo2020}.

In conclusion, a sensitivity for the $\text{Z} \rightarrow \mu\text{e}$ mode
at the $10^{-8}$ level should be safely within reach at FCC-ee. An independent method
for e/$\mu$ separation, as that provided by a powerful d$E/$d$x$ measurement,
could improve this sensitivity by one to two orders of magnitude
potentially all the way down to the $10^{-10}$ level.


\subsubsection*{%
\texorpdfstring{$\boldsymbol{\text{Z} \rightarrow \tau\mathrm{e}}$}{Ztotaue}
and 
\texorpdfstring{$\boldsymbol{\text{Z} \rightarrow \tau\mu}$}{Ztotaumu}
}

In e$^+$e$^-$ collisions, the pursuit for $\text{Z} \rightarrow \tau\text{e}\ (\tau\mu)$ decays amounts to a search for
events with
a \emph{clear tau decay} in one hemisphere recoiling against
a \emph{beam-momentum electron (muon)} in the other. This procedure was used at LEP, where sensitivities were at the $10^{-5}$ level.

\begin{figure}[tb]
  \begin{center}
    \includegraphics[width=0.45\textwidth]{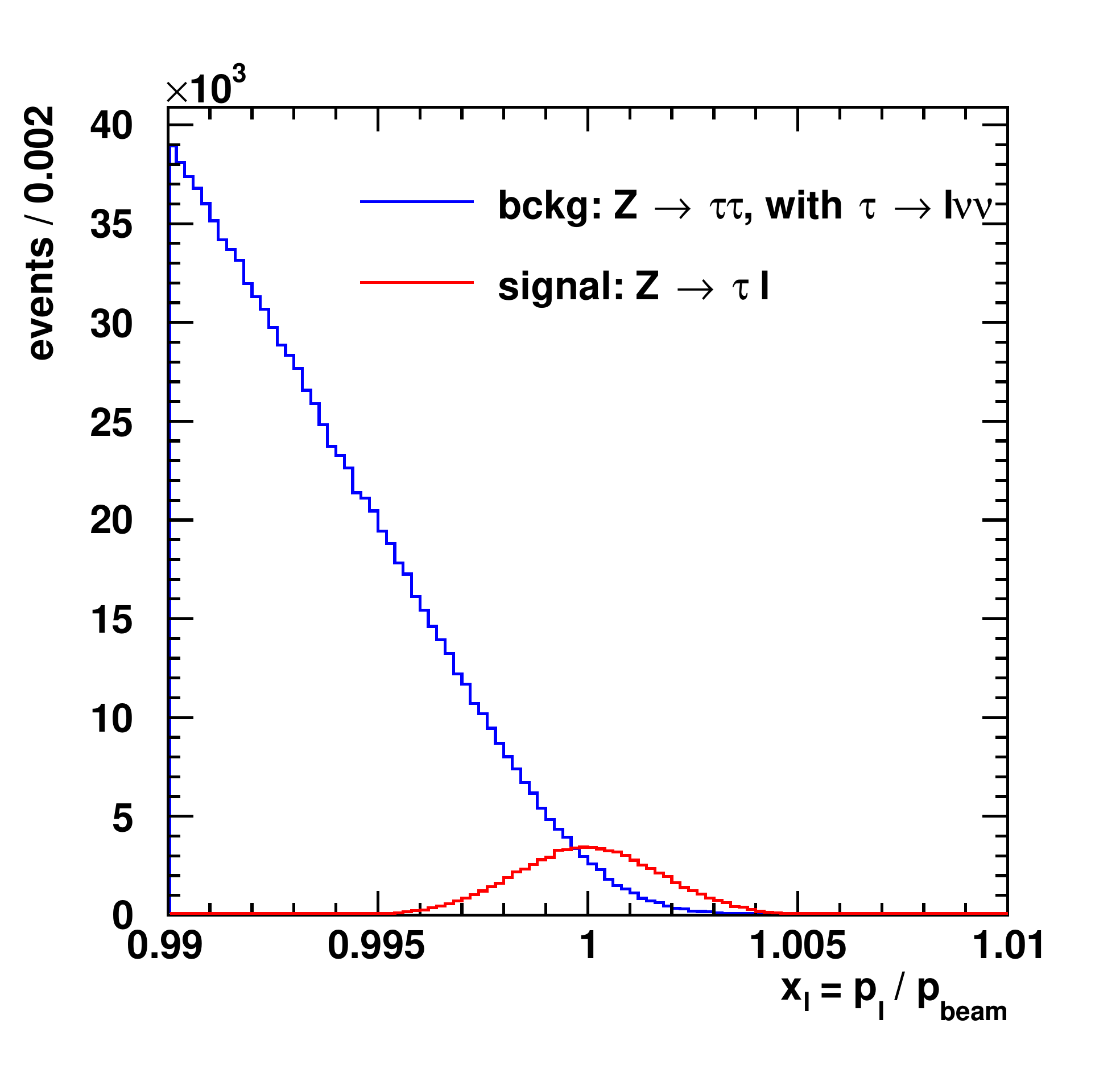}
    \caption{FCC-ee search for lepton flavour violating decays
      $\text{Z}\rightarrow\tau\ell$, $\ell=\mathrm{e},\mu$.
      Momentum distribution (upper part) of the final state lepton $\ell$ for the signal
      (red) and for the background from
      \mbox{$\text{Z}\rightarrow\tau\tau$}, with $\tau\rightarrow\ell\bar{\nu}\nu$
      (blue).
      A momentum resolution of $1.8 \times 10^{-3}$ has been assumed, see text.
      For illustration, the signal branching fraction is set here to
      \mbox{$\mathcal{B}(\text{Z}\to\tau\ell) = 10^{-7}$.}}
    \label{fig:ZLFV}
  \end{center}
\end{figure}

%
To illuminate the analysis procedure, 
first 
consider the term \emph{clear tau decay}. Here, the point is to restrict the
analysis to those decay modes, where the probability is minimal of
misidentifying a final-state lepton from $\text{Z}\to\mu\mu$
or $\text{Z}\to\text{ee}$ as a $\tau$ decay.
Immediately this excludes the
leptonic modes $\tau\to\mu\bar{\nu}\nu$ and $\tau\to\text{e}\bar{\nu}\nu$ and possible also the hadronic mode $\tau\to\pi\nu$, where the pion may be mis-identified as a lepton.
%
To reach very high purities, it may be ultimately necessary to
restrict the analysis to 
exclusive modes such as
$\tau\to\rho\nu\to\pi\pi^0\nu$ and modes with three (or more) charged particles.
Secondly, 
consider the term \emph{beam-momentum electron (muon)}. 
%
The momentum distribution of the final-state charged lepton is given by~\cite{tsaiTau1971}
%
\begin{equation}
  \frac{1}{\Gamma} \frac{\mathrm{d}\Gamma}{\mathrm{d}x} =
  \frac{1}{3} \left[\left(5-9x^2+4x^3\right) +
    P_\tau \left(1-9x^2+8x^3\right) \right],
\end{equation}
where $x=p/p_\mathrm{beam}$, and $P_\tau$ is the longitudinal polarisation of the
$\tau$ leptons. The density of events close to the beam-momentum endpoint depends on $P_\tau$,
which has the value $P_\tau \simeq -0.15$, as measured at LEP~\cite{EWWG2005}. 
The separation of signal and background now depends
on the experimental precision by which a \emph{beam-momentum particle} can be
defined. This is illustrated in Figure~\ref{fig:ZLFV}, where a momentum spread of $1.8\times 10^{-3}$ has been assumed. This value arises as a combination of the
$0.9\times 10^{-3}$ spread of the collision energy and the $1.5\times 10^{-3}$
momentum resolution typical for a FCC-ee detector at $p=45.6$~GeV.

A simple estimate of the search sensitivity was established by assuming a signal efficiency
of 25\% and by defining the signal box via the simple requirement $x>1$. 
For the quoted momentum spread, this procedure indicated a sensitivity to branching fractions down to $10^{-9}$.
The sensitivity was found to scale linearly in the momentum spread.

\subsection{Higgs Boson Decays}

The sensitivity of searches for CLFV Higgs decays, $\text{H}\to \text{e}\mu$, e$\tau$, and $\mu\tau$, at future e$^+$e$^-$ colliders has been estimated in \mbox{Ref.\ \cite{Qin:2017aju}}. The studies were based on the dominant Higgs production process $\mbox{e}^+\mbox{e}^- \to \mbox{ZH}$, with an assumed event multiplicity of about one million at circular colliders (CEPC and FCC-ee) and half a million at the ILC. Longitudinal beam polarisation was assumed at the ILC, leading to a somewhat larger HZ production cross section, but no other discernible effect. The dominant hadronic decay mode of the Z into quark pairs was used for all three Higgs decay modes, and relatively loose requirements on the di-jet invariant mass ensured a high reconstruction efficiency for the Z. 


For the $\text{H}\to \text{e}\mu$ mode, the electron--muon invariant mass was asked to be consistent, within a window, with the Higgs boson mass, leading to a signal detection efficiency of 41\%, and very few accepted background events. 
The upper bound on the $\text{H}\to \text{e}\mu$ branching fraction was found to be $\mathcal{B}(\text{H}\to \text{e}\mu) < 1.2 \times 10^{-5}$ ($2.1 \times 10^{-5}$) at CEPC and FCC-ee (ILC).

For the two 
modes involving $\tau$s, $\text{H}\to \text{e}\tau$ and $\text{H}\to\mu\tau$, the same event signature 
was used, \emph{i.e.}\ 
two jets, one electron, and one muon. With this choice, targeting the leptonic $\tau$ 
decay mode $\tau\rightarrow\mu\bar{\nu}\nu$ for $\text{H}\to \text{e}\tau$ ($\tau\rightarrow\text{e}\bar{\nu}\nu$ for $\text{H}\to \mu\tau$),
the $\tau$ was then reconstructed from the muon (electron) and the missing energy. 
With a ceiling on the reconstructed $\tau$ mass of a few GeV, the e$\tau$ ($\mu\tau$) invariant mass was then required to be consistent with the Higgs boson mass. 
The overall signal efficiency was at the 5\% level, nearly one order of magnitude lower than for the $\text{H}\to \text{e}\mu$ mode, with more than half of the difference caused by the $\sim$18\% $\tau$ leptonic branching fraction. The upper bound on the $\text{H}\to \text{e}\tau$ and $\text{H}\to \mu\tau$ branching fractions were found to be rather similar, so here we quote the mean of the two: $\mathcal{B}(\text{H}\to \ell\tau) < 1.5 \times 10^{-4}$ ($2.4 \times 10^{-4}$) at CEPC and FCC-ee (ILC).

\section{Summary}
\label{sec:conclusion}

The first precision results on CLFV decays of heavy particles came from LEP and were based on about $4 \times 10^6$ observed Z decays per experiment. Branching fraction limits were established at about $2 \times 10^{-6}$ for $\mbox{Z}\to \mbox{e}\mu$ and $10^{-5}$ for $\mbox{Z}\to \mbox{e}\tau,\,\mu\tau$. Recently, ATLAS has improved these limits by a factor of about seven for the $\mu$e mode and about two for the, obviously more difficult, $\tau$e and $\tau\mu$ modes. No results have so far been released by CMS. Sensitivities can be expected to improve by a factor $1/\sqrt{\mathcal{L}}$, where $\mathcal{L}$ is the collected luminosity, implying ultimately an improvement by an additional factor of about five from HL-LHC. Sensitivities would then be at the $10^{-7}$ level for $\mbox{Z}\to \mbox{e}\mu$ and at the $10^{-6}$ level for $\mbox{Z}\to \mbox{e}\tau,\,\mu\tau$. With its phenomenal Z-pole luminosity performance, FCC-ee will produce $5\times 10^{12}$ Z decays exceeding HL-LHC by about two orders of magnitude. Sensitivities to CLFV Z decays will depend strongly on detector performance. Of particular importance is the optimization of detectors in terms of momentum resolution and PID performance. Sensitivities could reach down to the $10^{-9}$ level. For the $\mu$e mode the sensitivity will depend critically on whether sufficient instrumental redundency is available to control precisely the fraction of muons appearing as electrons due to hard bremsstrahlung.

Searches for heavy resonant (and non-resonant) production of e$\mu$, e$\tau$, and $\mu\tau$ final states above the Z resonance have been reported by ATLAS and CMS. From their full Run-2 dataset, CMS set the most stringent limits, excluding a Z$'$ up to 5.0\,TeV in the e$\mu$ channel, 4.3\,TeV in the e$\tau$ channel, and 4.1\,TeV in the $\mu\tau$ channel.

Searches for LFV decays of the Higgs boson have been reported by ATLAS and CMS. The most stringent limits, extracted from the full Run-2 datasets, are $\mathcal{B}(H\to \mu\tau)<0.15\%$ and $\mathcal{B}(H\to e\tau)<0.22\%$, from CMS, and $\mathcal{B}(H\to e\mu)<6.1\times10^{-5}$, from ATLAS. Accounting for the 20 times larger HL-LHC datasets and possible improvements to the analysis, the modes involving taus could potentially reach sensitivities at the $1$--$5\times 10^{-4}$ level, whereas the e$\mu$ mode could reach the $10^{-5}$ level. Similar sensitivities have been estimated for CEPC and FCC-ee from their samples of about one million ZH events.

Complementary searches for the decay to $\mu\tau$ of a scalar particle has allowed LHCb and CMS to set branching fraction limits in a wide mass range.

From the more than $10^8$ top quarks produced during Run 2, CMS has established bounds on the $t \to u\mu$e and $t\to c\mu$e branching fractions at the level of $10^{-7}$ and $10^{-6}$, respectively. 
Improvements of about a factor five can be expected from HL-LHC. Future e$^+$e$^-$ colliders will be limited by the about 10$^6$ $t\bar{t}$ events they will produce and will not be able to improve significantly on the HL-LHC results.

\bibliographystyle{JHEP}
\bibliography{theBIB}

\end{document}